\documentclass[letterpaper,12pt]{article}
\usepackage[dvipdfmx]{graphicx}
\usepackage{amsfonts,amsthm,amsmath,amssymb,graphicx,float,mathtools}
\numberwithin{equation}{section}
\usepackage{subfigure}
\usepackage{color}
\usepackage{bbm} 
\usepackage{tensor} 
\usepackage{verbatim} 
\usepackage[
pagebackref=false,
colorlinks=true,
linkcolor=blue,
urlcolor=blue,
filecolor=black,
citecolor=red,
pdfstartview=FitV,
pdftitle={},
pdfauthor={},
pdfsubject={},
pdfkeywords={},
pdfpagemode=None,
bookmarksopen=true
]{hyperref}
\usepackage{ascmac}
\usepackage{ulem}
\usepackage{booktabs}
\usepackage[usenames,dvipsnames,svgnames,table,x11names]{xcolor}
\usepackage{cite}
\begin{document}

\newtheorem{definition}{Definition}[section]
\newcommand{\be}{\begin{equation}}
\newcommand{\ee}{\end{equation}}
\newcommand{\bea}{\begin{eqnarray}}
\newcommand{\eea}{\end{eqnarray}}
\newcommand{\LE}{\left[}
\newcommand{\R}{\right]}
\newcommand{\nn}{\nonumber}
\newcommand{\Tr}{\text{Tr}}
\newcommand{\N}{\mathcal{N}}
\newcommand{\G}{\Gamma}
\newcommand{\vf}{\varphi}
\newcommand{\LL}{\mathcal{L}}
\newcommand{\Op}{\mathcal{O}}
\newcommand{\HH}{\mathcal{H}}
\newcommand{\arctanh}{\text{arctanh}}
\newcommand{\up}{\uparrow}
\newcommand{\down}{\downarrow}
\newcommand{\ket}[1]{\left| #1 \right>}
\newcommand{\bra}[1]{\left< #1 \right|}
\newcommand{\ketbra}[1]{\left|#1\right>\left<#1\right|}
\newcommand{\rd}{\partial}
\newcommand{\de}{\partial}
\newcommand{\ba}{\begin{eqnarray}}
\newcommand{\ea}{\end{eqnarray}}
\newcommand{\db}{\bar{\partial}}
\newcommand{\we}{\wedge}
\newcommand{\ca}{\mathcal}
\newcommand{\lr}{\leftrightarrow}
\newcommand{\f}{\frac}
\newcommand{\s}{\sqrt}
\newcommand{\vp}{\varphi}
\newcommand{\hvp}{\hat{\varphi}}
\newcommand{\tvp}{\tilde{\varphi}}
\newcommand{\tp}{\tilde{\phi}}
\newcommand{\ti}{\tilde}
\newcommand{\ap}{\alpha}
\newcommand{\pr}{\propto}
\newcommand{\mb}{\mathbf}
\newcommand{\ddd}{\cdot\cdot\cdot}
\newcommand{\no}{\nonumber \\}
\newcommand{\la}{\langle}
\newcommand{\lb}{\rangle}
\newcommand{\ep}{\epsilon}
\def\we{\wedge}
\def\lr{\leftrightarrow}
\def\f {\frac}
\def\ti{\tilde}
\def\ap{\alpha}
\def\pr{\propto}
\def\mb{\mathbf}
\def\ddd{\cdot\cdot\cdot}
\def\no{\nonumber \\}
\def\la{\langle}
\def\lb{\rangle}
\def\ep{\epsilon}
\newcommand{\mcl}{\mathcal}
\def\g{\gamma}
\def\Tr{\text{tr}}
\def\MWCommentColor{ForestGreen}
\newcommand{\mw}[1]{{\color{\MWCommentColor} [MW: {#1}]}}

\begin{titlepage}
\thispagestyle{empty}

\begin{flushright}

\end{flushright}
\bigskip

\begin{center}
\noindent{\large \textbf{Hawking-Page and entanglement phase transition\\ in $2$d CFT on curved backgrounds}}\\
\vspace{2cm}

\vspace{1cm}
\renewcommand\thefootnote{\mbox{$\fnsymbol{footnote}$}}
Akihiro Miyata\footnote{a.miyata@ucas.ac.cn}${}^{1}$,
Masahiro Nozaki\footnote{mnozaki@ucas.ac.cn}${}^{1,2}$,
 Kotaro Tamaoka\footnote{tamaoka.kotaro@nihon-u.ac.jp}${}^{3}$ and Masataka Watanabe\footnote{max.washton@gmail.com}${}^{4}$\\

\vspace{1cm}
${}^{1}${\small \sl Kavli Institute for Theoretical Sciences, University of Chinese Academy of Sciences,
Beijing 100190, China}\\
${}^{2}${\small \sl RIKEN Interdisciplinary Theoretical and Mathematical Sciences (iTHEMS), \\Wako, Saitama 351-0198, Japan}\\
${}^{3}${\small \sl Department of Physics, College of Humanities and Sciences, Nihon University, \\Sakura-josui, Tokyo 156-8550, Japan}\\
${}^{4}${\small \sl Graduate School of Informatics, Nagoya University, Nagoya 464-8601, Japan}\\

\vskip 4em
\end{center}
\begin{abstract}
The thermodynamics and the entanglement properties of two-dimensional conformal field theories ($2$d CFTs) on curved backgrounds are studied.
By means of conformal mapping we study the equivalent system on flat space governed by the deformed Hamiltonian, which is a spatial integral of the Hamiltonian density modulated by an enveloping function.
Focusing on holographic CFTs, we observe Hawking-Page like phase transition for the thermal and the entanglement entropy as we vary the background metric.
We also compute the mutual information to study the information theoretic correlation between parts of the curved spacetime.
The gravity dual of 2d CFTs on curved background is also discussed.
\end{abstract}
\end{titlepage} 
\tableofcontents
\section{Introduction and Summary}

The analysis of phase transitions is an important step toward understanding strongly-coupled systems.
For holographic systems, in particular, the phases correspond to semi-classical saddle points on AdS spacetime, making it an ideal way to study the topology and the connectivity of space.
By using the Ryu-Takayanagi formula \cite{Ryu:2006bv,Ryu:2006ef}, one can, for example, measure the distance between two distant points on the boundary of {three-dimensional anti-de sitter space} (${\rm AdS}_3$) by computing the entanglement entropy of the corresponding {two-dimensional conformal field theory} ($2$d CFT).

One of the most common phase transitions in holographic CFTs is the Hawking-Page transition \cite{Hawking:1982dh}.
It originally refers to the change of dominant semi-classical gravity saddle-points from thermal gas to AdS black holes as we heat up the system \cite{Banados:1992wn,Witten:1998zw}.
In the dual CFT context, this corresponds to a first-order confinement-deconfinement phase transition \cite{Maldacena:1997re,Witten:1998qj,Witten:1998zw,Gubser:1998bc} and as a consequence, holographic CFTs behave qualitatively differently at low and high temperature.
They also have qualitatively different entanglement structures in those two parameter regions as seen from the Ryu-Takayanagi formula.

The Hawking-Page transition is, however, in some sense an idealised situation, where we heat up the system in a homogeneous way.
In realistic experiments{,} we generically expect temperature gradients and so it is an interesting question to ask what happens to the phase transition when there is such an inhomogeneity.
As we explain later, the temperature gradient can be effectively modeled by putting the CFT on curved space, so the more general interest should be in the phase and the entanglement structure of CFTs on a curved background.
In the context of AdS/CFT, it is also interesting to study what those phases correspond to in dual gravity.
Moreover, putting CFTs on the de Sitter background might be a useful first step towards understanding the matter sector of our universe which is de Sitter.

One way to analytically study CFTs on curved backgrounds (or more specifically with temperature gradients) is to restrict oneself to two dimensions.
We start from the Hamiltonian on flat space which is an integral over space of the original Hamiltonian density $\mathcal{H}(x)$ multiplied by the enveloping function $f(x)$.
In two dimensions, it is an integral on a complex plane $w$ with the contour on the imaginary axis running from $w=0$ to $w=2\pi i$,
\begin{equation}
    H_f=\int_{C} \frac{dw}{2\pi i} f(w) T(w)+\rm{(c.c.)},
\end{equation}
where $\Im w$ is the spatial direction and $\Re w$, the {Euclidean} time direction.
The enveloping function can be thought of as the local temperature or speed of light of the system.
Now, this flat space deformed Hamiltonian can equivalently be realised as a CFT on a curved background with metric
\begin{equation}
    ds^2=\left(f(x)dt\right)^2+dx^2,
\end{equation}
{ where $t$ is the Euclidean time.}
As complicated as it looks, the deformed Hamiltonian is simply a sum of Virasoro generators after radial quantization,
\begin{align}
    H_f=\sum_{n}f_nL_{n}+\rm{(const.)},
\end{align}
where $f(z)\equiv \sum_{n=-\infty}^{\infty} f_n z^n$
and hence it is analytically tractable.
The aim of the present paper is therefore to study the thermal and entanglement structure of such deformed Hamiltonians.

We will focus on a particular type of the enveloping function in this paper, called the M\"obius deformation,
\begin{align}
    f(ix)=1-\tanh{2\theta}\left(1-2\sin^2{\left(\f{q \pi x}{L}\right)}\right),
\end{align}
where $\theta>0$ is a free parameter. {The M\"obius deformed $2$d CFTs were introduced in \cite{Tada:2014kza,Ishibashi:2015jba,Ishibashi:2016bey, Okunishi_2016,2016PhRvB..93w5119W} as a one-parameter generalization of the sine-square deformation (SSD) in the context of numerical calculations of spin systems~\cite{Gendiar01102009,Gendiar01022010,Hikihara_2011,2011PhRvA..83e2118G,Shibata:2011jup,Shibata_2011,Maruyama_2011,Katsura:2011zyx,Katsura:2011ss,PhysRevB.86.041108,PhysRevB.87.115128}. 
Indeed, by taking the $\theta\rightarrow\infty$ limit, the M\"obius deformation reduces to SSD. (Refer to developments~\cite{PhysRevB.97.184309,2019JPhA...52X5401M,2021arXiv211214388G,PhysRevLett.118.260602,2018arXiv180500031W,2020PhRvX..10c1036F,Han_2020,2021PhRvR...3b3044W,2020arXiv201109491F,2021arXiv210910923W,PhysRevB.103.224303,PhysRevResearch.2.023085,Moosavi2021,PhysRevLett.122.020201,10.21468/SciPostPhys.3.3.019,2016arXiv161104591Z,2020PhRvR...2c3347R,HM,2018PhRvL.120u0604A,2019PhRvB..99j4308M,2022arXiv221100040W,2023arXiv230208009G,2023arXiv231019376N,2023arXiv230501019G,Das:2023xaw,2023arXiv230904665K,Liu:2023tiq,2024arXiv240315851M,2024arXiv240501642L,2024arXiv240216555B,2024arXiv240407884J} for further developments on the subject.)}
We will compute the entanglement entropy and the mutual information of the Gibbs state of the deformed Hamiltonian, \it i.e., \rm $\rho=\f{e^{-\beta H_{f}}}{\Tr e^{-\beta H_{f}}}$, at various temperature, to detect the Hawking-Page like phase transition.


\subsection*{Summary of results}

In this paper, we explore the global (thermodynamic) and local (entanglement) properties of the thermal state on curved spacetime in $2$d holographic CFTs.
We have two main results: 
\begin{itemize}
    \item[] {\bf Thermodynamic property:} 
    We find the first-order phase transition by varying the M\"obius background metric using $\theta$ by computing the thermal entropy of the deformed Hamiltonian.
    We point out that this is because the Hamiltonian on curved space is conformally equivalent to the flat space Hamiltonian with size given by $L\cosh (2\theta)$.
    \item[] {\bf Entanglement property:} 
    We compute the entanglement entropy on the M\"obius background at finite temperature.
    We again find the first-order phase transitions, but only when the subsystem contains points where $\sin^2\left(q\pi x/L\right)=0$. 
    Otherwise no phase transitions occur for the entanglement entropy.
    We summarise the behaviour of the entanglement entropy at various temperature and the parameter $\theta$ in table \ref{table:ee-in-various}.
\end{itemize}

\begin{table}[t]
 \label{table:ee-in-various}
 \label{table:SpeedOfLight}
 \centering
  \begin{tabular}{|c|c|c|c|c|c|}
   \hline
   & $\f{L \cosh{2\theta}}{\beta}<1, \theta \gg 1$ &\multicolumn{2}{|c|}{$\f{L \cosh{2\theta}}{\beta}>1, \theta \gg 1$}  &  \multicolumn{2}{|c|}{$\f{L}{\beta}\gg 1$}\\
   \hline 
    &   & $\f{L}{\beta} \ll 1$ &$\f{L}{\beta} \gg 1$ & $\theta \ll 1$& $\theta \gg 1$\\
    \hline 
   $S_{A_1}$& $ \f{c}{3}\log{\left[\f{L}{q\pi}\sin{\left(\f{q \pi l_{A_1}}{L}\right)}\right]}$  & $ \f{c}{3}\log{\left[\f{L}{q\pi}\sin{\left(\f{q \pi l_{A_1}}{L}\right)}\right]}$& $ \propto \f{L}{\beta}$& $\propto \f{l_{A_1}}{\beta}$&  $\propto \f{L}{\beta}$\\
   \hline
    $S_{A_2}$& $\propto \theta$  & \multicolumn{2}{|c|}{$\propto \f{L e^{2\theta}}{\beta}$} & $\propto \f{l_{A_2}}{\beta}$ & $\propto \f{L e^{2\theta}}{\beta}$\\
   \hline
  \end{tabular}
   \caption{The finite temperature entanglement entropy of the M\"obius Hamiltonian for various intervals of length $l_{A_{1,2}}$. $A_{1}$ ($A_{2}$) denotes a class of subsystems which contains (does not contain) points defined by $\sin^2\left(q\pi x/L\right)=0$.}
\end{table}

We furthermore computed the $\theta$ dependence of the mutual information at finite temperature between various pairs of intervals.
We also studied the gravity dual of the finite temperature system of our interest, which was determined to be the Black Hole with inhomogenous horizon radius.



\subsection*{Organization}
The organization of the rest of the paper is as follows. In Section \ref{sec:cuvature}, we introduce the deformed thermal Gibbs state by M\"obius Hamiltonian and discuss curvature dependence of the corresponding spacetime. In Section \ref{sec:Tharmal-and-EE}, we discuss details of the phase structure for the deformed thermal Gibbs state. We study global phase and thermal entropy in Section \ref{subsec:thermal-entropy}. In the rest of the subsection, we investigate local phase structure via entanglement entropy and mutual information. In Section \ref{sec:gravitydual}, we study the gravity dual of the deformed thermal Gibbs state. We conclude with some discussion in Section \ref{sec:discussions}.

\section{Thermal state on the curved background}\label{sec:cuvature}
\begin{figure}[ht]
\begin{center}
\begin{tabular}{c}
\subfigure[The $\theta$ dependence of the curvature; $q=1$.]{
\includegraphics[scale=0.78]{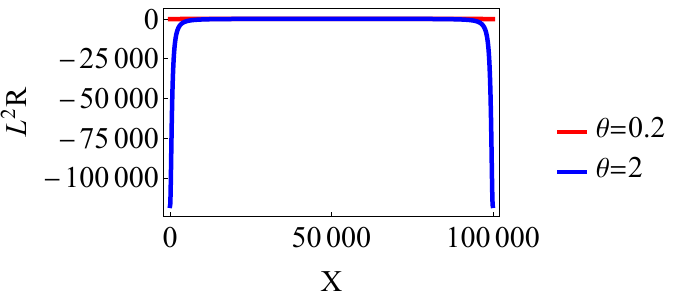}
\label{left1}
} \\
\subfigure[The $q$ dependence of the curvature; $\theta=0.5$. ]{
\includegraphics[scale=0.78]{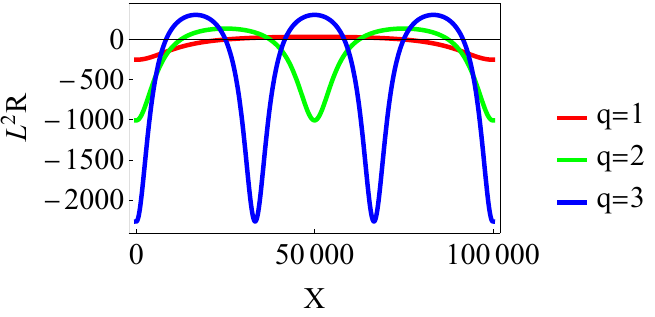}
\label{left2}
}
\end{tabular}
 \centering
\caption{Plots of the curvature (\ref{eq:RicciCurvature}) times the square of the period scale, $L^{2}$. The period is set to be $L=100000$. In (a), we show the {$\theta$ dependence} of the curvature, while in (b), we show the $q$-dependence of the curvature. The curvature can be either positive, zero, or negative depending on the location $x$.} 
\label{fig:curvature}
\end{center}
\end{figure}

We begin by defining the thermal state on the curved background as,
\be \label{eq:thermal-state-with-inhomogeneous}
\rho=\f{e^{-\beta H_{q-\text{M\"obius}}}}{\Tr e^{-\beta H_{q-\text{M\"obius}}}},
\ee
where $\beta$ is the inverse temperature, and the inhomogeneous Hamiltonian in two-dimensional conformal field theories ($2$d CFTs), $H_{q-\text{M\"obius}}$, is defined as 
\be \label{eq:H-q-mobius}
H_{q-\text{M\"obius}} = \int^L_0dx \left[1-\tanh{2\theta}\left(1-2\sin^2{\left(\f{q \pi x}{L}\right)}\right)\right](T(x)+\overline{T}(x)),
\ee
where the system is on the spatial circle with the circumstance of $L$, $\theta$ is a {non-negative} real parameter, and $q$ is a positive integer.
The Hamiltonian density is modulated by an envelop function defined by
\be\label{eq:enveFunc}
f(x,\theta)=1-\tanh{2\theta}\left(1-2\sin^2{\left(\f{q \pi x}{L}\right)}\right).
\ee
Since $f(x,\theta=0)=1$, $H_{q-\text{M\"obius}} $ with $\theta=0$ reduces to uniform Hamiltonian,
\be
H=\int^L_0dx (T(x)+\overline{T}(x))=\f{2\pi}{L}\left[\mathcal{L}^{z}_0+\overline{\mathcal{L}}^{\overline{z}}_0-\f{c}{12}\right],
\ee
where $c$ is the central charge, and the Virasoro's generators, $\mathcal{L}^{z}_0$ and $\overline{\mathcal{L}}^{z}_0$, are defined as 
\be
\mathcal{L}^{z}_n = \oint \f{dz}{2\pi i}z^{n+1}T(z),~ \overline{\mathcal{L}}^{\overline{z}}_n = \oint \f{d\overline{z}}{2\pi i}\overline{z}^{n+1}\overline{T}(\overline{z}).
\ee
Here, $n$ is an integer, $(z,\overline{z})$ is defined as $(z,\overline{z})=\left(e^{\f{2\pi (ix+\mathcal{T})}{L}}, e^{\f{2\pi (-ix+\mathcal{T})}{L}}\right)$, where $x, \mathcal{T} \in R$, and $T$ and $\overline{T}$ are holomorphic and anti-holomorphic pieces of the energy-momentum tensor.
The inhomogeneous Hamiltonian in (\ref{eq:H-q-mobius}) is equivalent to the uniform Hamiltonian on the curved background,
\be\label{eq:background}
\begin{split}
    &H_{q-\text{M\"obius}}=\int^{L}_0 {dx} \sqrt{-\det{g(x)}} (T(x)+\overline{T}(x)),\\
    &ds^2 =g_{ab}(x)dx^{a}dx^b= -f^2(x,\theta)dt^2+dx^2,
\end{split}
\ee
where $\theta$ is the parameter determining the geometrical structure of the background. 
As in \cite{Caputa:2020mgb} curvature of this geometry is given by
\be \label{eq:RicciCurvature}
R=-\f{2\partial^2_xf(x,\theta)}{f(x,\theta)}=\frac{8 \pi ^2 q^2 \tanh (2 \theta ) \cos \left(\frac{2 \pi  q x}{L}\right)}{L^2 \left(\tanh (2 \theta ) \cos \left(\frac{2 \pi  q x}{L}\right)-1\right)}.
\ee
Thus, the curvature depends on the spatial location $x$ and the parameter $\theta$.
In other words, (\ref{eq:thermal-state-with-inhomogeneous}) can be considered as the $2$d CFT Hamiltonian on the background (\ref{eq:background}).
Here, let us check the behavior of the curvature. First, some plots of the curvature as a function of the position $x$ are given in Fig. \ref{fig:curvature}.

From Fig. \ref{fig:curvature}, we can see that the curvature \eqref{eq:RicciCurvature} can be either positive, zero, or negative depending on the location $x$. In particular, the curvature \eqref{eq:RicciCurvature} vanishes at 
\be
x=\frac{L}{4q} +\frac{mL}{q}, \quad \frac{3L}{4q} +\frac{mL}{q} \qquad (m\in \mathbb{Z}),
\ee
and it becomes\footnote{The $x$-derivative of the curvature is given by 
\be
\partial_{x} R= \frac{2 \partial_{x}f(x,\theta ) \partial_{x}^{2}f(x,\theta )-2 f(x,\theta ) \partial_{x}^{3}f(x,\theta )}{f(x,\theta )^2} = \frac{16 \pi ^3 q^3 \tanh (2 \theta ) \sin \left(\frac{2 \pi  q x}{L}\right)}{L^3 \left(\tanh (2 \theta ) \cos \left(\frac{2 \pi  q x}{L}\right)-1\right)^2},
\ee
thus the zeros of the derivative are mainly determined by the numerator. One can imagine that the numerator is related to the derivative of the envelop function, which is given by
\be
\partial_{x}f(x,\theta )=\frac{2 \pi  q \tanh (2 \theta ) \sin \left(\frac{2 \pi  q x}{L}\right)}{L}.
\ee
} maximum
\be
\frac{8 \pi ^2 q^2 \tanh (2 \theta )}{L^2 (\tanh (2 \theta )+1)}>0 \quad \text{ at } x= \frac{L}{2q} +\frac{mL}{q} \qquad (m\in \mathbb{Z}),
\ee
and minimum
\be \label{eq:curvature-min}
\frac{8 \pi ^2 q^2 \tanh (2 \theta )}{L^2 (\tanh (2 \theta )-1)} <0 \quad \text{ at } x= \frac{mL}{q} \qquad (m\in \mathbb{Z}).
\ee
Also, the curvature becomes positive and negative for the intervals 
\be
R=
\begin{dcases}
    R&>0 \quad  \text{ for } \quad  \frac{L}{4q} +\frac{mL}{q} < x < \frac{3L}{4q} +\frac{mL}{q},\\
    R&<0  \quad  \text{ for } \quad  \frac{3L}{4q} +\frac{mL}{q} < x < \frac{L}{4q} +\frac{(m+1)L}{q}.
\end{dcases}
\qquad (m\in \mathbb{Z}).
\ee
Note that, in the SSD limit $\theta \to \infty$, the minimum of the curvature diverges as $-\frac{4 \pi^2 q^2}{L^2} e^{4 \theta } \to - \infty$, but the maximum of that remains finite $\frac{4 \pi ^2 q^2}{L^2}$. 

To clarify the property of the thermal state in (\ref{eq:thermal-state-with-inhomogeneous}), define the complex coordinates by 
\be
\begin{split} \label{eq:xi}
   & \xi=\frac{L_{\text{eff}}}{2\pi }\log\left[\chi\right],~ \overline{\xi}=\frac{L_{\text{eff}}}{2\pi }\log\left[\overline{\chi}\right],~\chi=\left(\frac{\cosh(\theta)z^q-\sinh(\theta)}{\cosh(\theta)-\sinh(\theta)z^q}\right)^{\f{1}{q}},~\overline{\chi}=\left(\frac{\cosh(\theta)\overline{z}^q-\sinh(\theta)}{\cosh(\theta)-\sinh(\theta)\overline{z}^q}\right)^{\f{1}{q}},
\end{split}
\ee
where $(z,\overline{z})=(e^{\f{2\pi w}{L}},e^{\f{2\pi \overline{w}}{L}})$, $(w,\overline{w})=(T+ix,T-ix)$, and $L_{\text{eff}}=L \cosh{2\theta}$.
By using the Virasoro's generators defined on the appropriate complex coordinate, $H_{q-\text{M\"obius}}$ is can be expressed by
\be \label{eq:H-q-mobius-virasoro}
H_{\text{q-M\"obius}}=\f{2\pi }{L_{\text{eff}}}\left[L^{\chi}_0+\overline{L}^{\overline{\chi}}_0-\f{c}{12}\right]+\f{2\pi cq^2}{12 L_{\text{eff}}}-\f{2\pi cq^2}{12 L}.
\ee
Consequently, (\ref{eq:thermal-state-with-inhomogeneous}) can be expressed by
\be \label{eq:themalstate}
\rho=\f{\exp{\left(\f{-2\pi  \beta\left(L^{\chi}_0+\overline{L}^{\overline{\chi}}_0\right)}{L_{\text{eff}}}\right)}}{\Tr \exp{\left(\f{-2\pi  \beta\left(L^{\chi}_0+\overline{L}^{\overline{\chi}}_0\right)}{L_{\text{eff}}}\right)}}.
\ee
This thermal state is given by replacing $\beta/L$ of the uniform thermal state, $e^{-\beta H}/\Tr e^{-\beta H}$, with $\beta/L\cosh{2\theta}$.
\section{Global and local properties of the thermal state on curved background \label{sec:Tharmal-and-EE}}

In this section, we will explore the global (thermodynamical) and local (entanglement) properties of the thermal state on the curved background in $2$d holographic CFTs.

\subsection{Thermal entropy}\label{subsec:thermal-entropy}
Now, we explore the thermodynamical properties of the thermal state in (\ref{eq:thermal-state-with-inhomogeneous}).
As in \cite{Witten:1998zw}, the thermodynamical property of the thermal partition function is determined by the moduli parameter defined as $\tau_{\text{Mod.}}=\f{L_{\text{eff}}}{ \beta}$\footnote{{For $2$d CFT on locally flat space, we usually discuss the phase transition by changing the ratio $L/\beta$. 
Here, we fix the ratio $L/\beta$ and change the ratio $L_{\text{eff}}/\beta$ by changing the parameter $\theta$. }}.
For $\tau_{\text{Mod.}}>1$, the behavior of the thermal partition function is described by the BTZ black hole, while for $\tau_{\text{Mod.}}<1$, that one is described by the thermal AdS$_3$.
The $\tau_{\text{Mod.}}$-dependence of the thermal partition function on the curved geometry is equivalent to that on the flat geometry with replacement of $L/\beta$ with $\tau_{\text{Mod.}}$ \footnote{The details of geometry is presented in Section \ref{sec:gravitydual}.}.
Therefore, the {$\theta$ dependence} of the thermal entropy is given by
\be
\begin{split}
    S_{\text{Thermal}} =\begin{cases}
    \mathcal{O}(1)~&~ \text{for} ~\tau_{\text{Mod.}}<1\\
        \f{c \pi L_{\text{eff}}}{3 \beta}~&~ \text{for} ~\tau_{\text{Mod.}}>1
    \end{cases}, 
\end{split}
\ee
where $\mathcal{O}(1)$ means that the leading term in the large $c$ expansion is at the order of one.
Furthermore, we determine the critical $\theta$, where the system exhibits the first-order phase transition.
The critical point, $\theta_c$, is given by
\be\label{eq:critical-theta}
\begin{split}
    \theta_{c}=\f{1}{2}\cosh^{-1}{\left(\f{\beta}{L}\right)} \underset{L/\beta \ll 1} {\approx} \f{1}{2}\log{\left(\f{2\beta}{L}\right)}
\end{split}
\ee
where $\cosh^{-1}$ denotes the inverse function of $\cosh$.
Thus, even for the low temperature regime, $L/\beta \ll 1$, the system can exhibit the phase transition.

\subsection{Entanglement entropy}
Next, we explore the entanglement property of the thermal system on the curved geometry by using entanglement entropy.

We define two intervals on which we compute the entanglement entropy.
Let $A_1$ be an interval $\left[X_1,X_2\right]$ which satisfies $\f{(n+1)L}{q}>X_1>X_2>\f{nL}{q}$, while $A_2$ an interval $\left[X_1,X_2\right]$ which satisfies $\f{(n+l+1)L}{q}>X_1>\f{(n+l)L}{q},~\f{(n+1)L}{q}>X_1>\f{nL}{q}$ where $n$ and $l$ are non-negative integers.
In other words, $A_1$ includes points on which $\sin^2\left(q\pi x/L\right)=0$, while $A_2$ does not.


Now the finite-temperature $n$-th R\"enyi entanglement entropy of our system can be computed using the twist operators as \cite{2009JPhA...42X4005C,2004JSMTE..06..002C}
\be
\begin{split}
    S^{(n)}_{A_i}&= \f{1}{1-n}\log{\left[\f{\Tr \sigma_{n}(w_1,\overline{w}_1) \overline{\sigma}_{n}(w_2,\overline{w}_2)e^{-\beta H_{\text{$q$-M\"obius}}}}{\Tr e^{-\beta H_{\text{$q$-M\"obius}}}}\right]}\\
    &=\f{1}{1-n}\log{\left[\prod_{i=1,2}\left(\f{d\xi_i}{dw_i}\right)^{h_n}\left(\f{d\bar{\xi}_i}{d\bar{w}_i}\right)^{\bar{h}_n}\right]}+\f{1}{1-n}\log{\left\langle \sigma_{n}(\xi_1,\overline{\xi}_1) \overline{\sigma}_{n}(\xi_2,\overline{\xi}_2)\right\rangle_{\text{Torus}}},
\end{split}
\ee
where the second line equality implements the conformal transformation.
Here $\sigma_n$ and $\bar\sigma_n$ is the twist and anti-twist operators with the conformal dimensions, $\left(h_{n},\bar{h}_{n}\right)=\left(\f{c(n^2-1)}{24n},\f{c(n^2-1)}{24n}\right)$ and $\left\langle \cdot \right \rangle_{\text{Torus}}$ means the expectation value on the torus with $L_{\text{eff}}$ and $\beta$, the spatial and Euclidean temporal periods.
Note that the result reduces to a trivial equality when $f(x)$ is a constant.

In the von Neumann limit, $n\rightarrow 1$, the R\'enyi entanglement entropy reduces to 
\be
\begin{split}\label{eq:EEU+NU}
    S_{A_i}=\f{c}{6}\log{\left[\prod_{i=1,2}\left(1-\tanh{2\theta}\cos{\left(\f{2q \pi X_i}{L}\right)}\right)\right]}+S^{\text{T.D.}}_{A_i}.
\end{split}
\ee
where $S^{\text{T.D.}}_{A_i}$ is the finite-temperature entanglement entropy on a circle of radius $L_{\rm eff}$, at inverse temperature $\beta$.
T.D. stands for ``theory dependent'', reflecting the fact that the first term is simply a conformal factor and is theory independent, while the latter is not.

It is obvious that the second term $S^{\rm T.D.}$ experiences the first-order phase transition at $\tau_{\text{Mod.}}\equiv L_{\rm eff}/\beta=1$ because of the Hawking-Page transition \cite{Barrella:2013wja}.
We will hereafter call $\tau_{\text{Mod.}}$ the effective temperature.
The cases $\tau_{\text{Mod.}}<1$ and $\tau_{\text{Mod.}}>1$ will thus be treated separately.


Hereafter we will only consider holographic CFTs, so understand that the equality is only at leading order at large central charge, $c$.

\subsubsection{Low effective temperature region $\tau_{\text{Mod.}}<1$}

First, we consider the {$\theta$ dependence} of $S_{A_{i=1,2}}$ in the low effective temperature region, $\tau_{\text{Mod.}}<1$.
This necessitates that $L/\beta<1$, too.
In the low effective temperature region, the theory-dependent piece of the entanglement entropy in $2$d holographic CFT is given by \cite{Barrella:2013wja}
\be
\begin{split}
    S^{\text{T.D.}}_{A_i}= \f{c}{3}\log{\left(\f{L_{\text{eff}}}{\pi\epsilon}\right)} +\f{c}{6}\log{\left|\sinh{\left(\f{\pi (\xi_1-\xi_2)}{L_{\text{eff}}}\right)}\right|^2} ,
\end{split}
\ee
with $\epsilon$ the UV cutoff.
This comes from the area of the minimal surface in the thermal AdS background. 
Introducing the coordinate $\varphi=-\f{i}{2}\log\chi$, the above expression can be written as
\be\label{eq:EEvac}
\begin{split}
    S_{A_i}^{\text{T.D.}}= \f{c}{3}\log{\left(\f{L_{\text{eff}}}{\pi\epsilon}\right)}+\f{c}{3}\log{\sin{\left[\f{\varphi_1-\varphi_2}{q}\right]}},
\end{split}
\ee
where $\varphi_{i}$ is given by\footnote{{In the $\theta \to 0$ limit, $\varphi_{i}$ is just given by $\varphi_{i}=\frac{q\pi}{L}X_{i}.$ }}
\begin{align}\label{eq:angle}
e^{i\varphi_{i}}& =\f{ e^{-\theta} \cos\left(\f{q \pi X_{i}}{L}\right) + i e^{\theta} \sin\left(\f{q \pi X_{i}}{L}\right) }{ \sqrt{ e^{-2\theta} \cos^{2}\left(\f{q \pi X_{i}}{L}\right) + e^{2\theta} \sin^{2}\left(\f{q \pi X_{i}}{L}\right) }  } \notag \\ 
	&=i e^{m_{i}\pi i } \f{1-i e^{-2\theta} \cot\left(\f{q \pi X_{i}}{L}\right)}{ \sqrt{1+ e^{-4\theta} \cot^{2}\left(\f{q \pi X_{i}}{L}\right)}} \quad \left( e^{m_{i}\pi i }=\f{\sin\left(\f{q \pi X_{i}}{L}\right)}{|\sin\left(\f{q \pi X_{i}}{L}\right)|} \right). 
\end{align}


Let us now focus on the large $\theta$ limit, where we can further simplify the computation as well as probe the most inhomogeneous region in terms of the enveloping function.
This limit necessarily leads to $L/\beta\ll 1$.

\subsubsection*{Large $\theta$ limit}
In the {large $\theta$} limit, by using the expression \eqref{eq:angle}, we can approximate the angle  $\varphi$ by
\be\label{eq:compPhiExpan}
\begin{split}
	e^{i\varphi_{i}}\underset{\theta \gg 1}{\approx}& ie^{m_{i}\pi i }\Bigg[ 1 -i e^{-2\theta} \cot\left(\f{q \pi X_{i}}{L}\right) -\f{1}{2}e^{-4\theta}  \cot^{2}\left(\f{q \pi X_{i}}{L}\right)+ \mathcal{O}(e^{-6\theta})\Bigg].
\end{split}
\ee
Also, more directly, we can express the angle $\varphi$ as
\begin{equation}\label{eq:AngleExpanLargeTheta}
	\begin{aligned}
		\varphi_{i} &\underset{\theta \gg 1}{\approx} m_{i} \pi + \frac{1}{2}\pi-i \log  \Bigg[ 1 -i e^{-2\theta} \cot\left(\f{q \pi X_{i}}{L}\right) -\f{1}{2}e^{-4\theta}  \cot^{2}\left(\f{q \pi X_{i}}{L}\right)+ \mathcal{O}(e^{-6\theta})\Bigg] \\
	 & \underset{\theta \gg 1}{\approx} m_{i} \pi + \frac{1}{2}\pi - e^{-2\theta} \cot\left(\f{q \pi X_{i}}{L}\right)  + \mathcal{O}(e^{-6\theta}).
	\end{aligned}
\end{equation}
We note that this approximation for $\varphi$ is valid only for regions where the infinitesimal parameter $ e^{-2\theta} $ is sufficiently smaller than $\cot\left(\f{q \pi X_{i}}{L}\right)$; $e^{-2\theta}\cot\left(\f{q \pi X_{i}}{L}\right)  \ll 1$.
Then, $\sin{\left[\f{\varphi_1-\varphi_2}{q}\right]}$ is approximated by
\be\label{eq:angleSin}
\sin{\left[\f{\varphi_1-\varphi_2}{q}\right]} \underset{\theta \gg 1}{\approx}
\begin{dcases}
	\f{e^{-2\theta}}{q} \cdot  \frac{\sin \left[\frac{q \pi\left(\hat{X}_1-\hat{X}_2\right)}{L}\right]}{\sin \left[\frac{q \pi \hat{X}_1}{L}\right] \sin \left[\frac{q \pi \hat{X}_2}{L}\right]} & m_{1}-m_{2}=0\\
	\sin\left( \f{l}{q}\pi\right) +  \f{e^{-2\theta}}{q} \cdot  \frac{\sin \left[\frac{q \pi\left(\hat{X}_1-\hat{X}_2\right)}{L}\right]}{\sin \left[\frac{q \pi \hat{X}_1}{L}\right] \sin \left[\frac{q \pi \hat{X}_2}{L}\right]} \cos\left( \f{l}{q}\pi\right) & m_{1}-m_{2}=l\\
\end{dcases},
\ee
where $\hat{X}_i$ is defined by $X_i=\f{m_{i}L}{q}+\hat{X}_i$.
Consequently, the theory-dependent pieces for the two cases are given at large $\theta$ by
\be\label{eq:nonUniEELow}
\begin{split}
	&S^{\text{T.D.}}_{A_1} \approx   \f{c}{3}\log{\left(\f{L e^{2\theta}}{2\pi\epsilon}\right)}+\f{c}{3}\log{ \left[\f{e^{-2\theta}}{q} \cdot  \frac{\sin \left[\frac{q \pi\left(\hat{X}_1-\hat{X}_2\right)}{L}\right]}{\sin \left[\frac{q \pi \hat{X}_1}{L}\right] \sin \left[\frac{q \pi \hat{X}_2}{L}\right]} \right] }\\
	&S^{\text{T.D.}}_{A_2} \approx   \f{c}{3}\log{\left(\f{L e^{2\theta}}{2\pi\epsilon}\right)}+\f{c}{3}\log{ \sin\left( \f{l}{q}\pi\right)  }.
\end{split}
\ee
Combining these theory-dependent pieces and the universal one, we obtain
\be 
 \begin{split}\label{eq:EE-Low-LargeTheta}
 	S_{A_1} &\approx \f{c}{6}\log{\left[4\prod_{i=1,2}\sin^2{\left(\f{q\pi X_i}{L}\right)}\right]} + \f{c}{3}\log{\left(\f{L e^{2\theta}}{2\pi\epsilon}\right)}+\f{c}{3}\log{ \left[\f{e^{-2\theta}}{q} \cdot  \frac{\sin \left[\frac{q \pi\left(\hat{X}_1-\hat{X}_2\right)}{L}\right]}{\sin \left[\frac{q \pi \hat{X}_1}{L}\right] \sin \left[\frac{q \pi \hat{X}_2}{L}\right]} \right] } \\
 	&=  \f{c}{3} \log \left[ \f{L}{\pi q \epsilon} \sin \left[\frac{q \pi\left(\hat{X}_1-\hat{X}_2\right)}{L}\right]  \right]\\
 	S_{A_2} &\approx  \f{c}{3}\log{\left(\f{L e^{2\theta}}{2\pi\epsilon}\right)} = \f{2c}{3} \theta + \f{c}{3}\log{\left(\f{L}{2\pi\epsilon}\right)}.
 \end{split}
\ee
We note that these expressions are obtained under the condition $e^{-2\theta}\cot\left(\f{q \pi X_{i}}{L}\right) {\ll} 1 $. 
In other words, we are taking the entangling points to be $O(1)$ away from the points defined by $\sin^2(q\pi X/L)=0$.
In the effective low temperature region with $\theta \gg 1$, $S_{A_1}$ approximately reduces to the vacuum entanglement entropy in the system with the size of $L/q$, while $S_{A_2}$ is independent of the subsystem size, and linearly grows with $\theta$. 
\subsubsection{High effective temperature region  $\tau_{\text{Mod.}}>1$}
We will explore the {$\theta$ dependence} of $S_{A_{i=1,2}}$ in the high effective temperature region, $\tau_{\text{Mod.}}>1$.
This can be separated into two regions, $L/\beta>1$ or $L/\beta<1$, which we discss later. 
In the high effective temperature region, the theory-dependent piece in $2$d holographic CFT, $S^{\text{T.D.}}_{A_i}$, is determined by \cite{Barrella:2013wja}
\be\label{eq:theEEOrigi}
\begin{split}
    S^{\text{T.D.}}_{A_i}&= \f{c}{3}\log{\left(\f{\beta}{\pi\epsilon}\right)} \\
    &\quad+\text{Min}\left[\f{c}{6}\log{\left|\sin{\left(\f{\pi (\xi_1-\xi_2)}{\beta}\right)}\right|^2}, \f{c}{6}\log{\left|\sin{\left(\f{\pi\left(\pm iL_{\text{eff}}- (\xi_1-\xi_2)\right)}{\beta}\right)}\right|^2}+\f{c\pi L_{\text{eff}}}{3\beta}\right],
\end{split}
\ee
This comes from the area of the minimal surface in the BTZ background, where we choose the smaller one by comparing the connected and the disconnected RT surface \cite{Rangamani:2016dms}.
Pluggin in the expressions for $\xi$, this becomes
\be\label{eq:theEE}
\begin{split}
    S_{A_i}^{\text{T.D.}}&= \f{c}{3}\log{\left(\f{\beta}{\pi\epsilon}\right)}\\
    &\quad+\text{Min} \left[\f{c}{3}\log{\left[\sinh{\left[\f{L_{\text{eff}}\left(\varphi_1-\varphi_2\right)}{q\beta}\right]}\right]}, \f{c}{3}\log{\left|\sinh{\left[\f{L_{\text{eff}}\left(\varphi_1-\varphi_2 \mp \pi q \right)}{q\beta}\right]}\right|}+\f{c\pi L_{\text{eff}}}{3\beta}  \right],
\end{split}
\ee
where $\varphi_i$ is given by \eqref{eq:angle}.


We hereby focus on two regimes of interest, $\theta\gg 1$ and $L/\beta\gg 1$. 
For the former we will obtain the analytic form of the entanglement entropy at large and small $L/\beta$, while for the latter we do so at large and small $\theta$.

\subsubsection*{Large $\theta$ limit}
At leading order in the large $\theta$ limit, $S^{\text{T.D.}}_{A_i}$ is given by
\be\label{eq:EEforHigh-LargeTheta}
\begin{split}
    &S^{\text{T.D.}}_{A_1} \approx \f{c}{3}\log{\left(\f{\beta}{\pi\epsilon}\right)}+\f{c}{3}\log{\left[\sinh{\left[\f{L\sin{\left[\f{q\pi (\hat{X}_1-\hat{X}_2)}{L}\right]}}{2q\beta \sin{\left[\f{q\pi \hat{X}_1}{L}\right]}\sin{\left[\f{q\pi \hat{X}_2}{L}\right]}}\right]}\right]},\\
    &S^{\text{T.D.}}_{A_2} \approx \f{c}{3}\log{\left(\f{\beta}{\pi\epsilon}\right)}+\f{c \pi Le^{2\theta}}{6 \beta}\cdot\f{l}{q},\\
\end{split}
\ee
where $\hat{X}_i$ is given by $X_i=\f{m_{i}L}{q}+\hat{X}_i$.
As a consequence, the entanglement entropies for $A_i$ are approximately given by
\be \label{eq:high-temperature-region-in-large-theta}
\begin{split}
    S_{A_1} &\approx \f{c}{6}\log{\left[4\prod_{i=1,2}\sin^2{\left(\f{q\pi X_i}{L}\right)}\right]}+\f{c}{3}\log{\left(\f{\beta}{\pi\epsilon}\right)}+\f{c}{3}\log{\left[\sinh{\left[\f{L\sin{\left[\f{q\pi (\hat{X}_1-\hat{X}_2)}{L}\right]}}{2q\beta \sin{\left[\f{q\pi \hat{X}_1}{L}\right]}\sin{\left[\f{q\pi \hat{X}_2}{L}\right]}}\right]}\right]}\\
   &\approx \begin{cases}
       \f{c}{3}\log{\left[\f{L \sin{\left[\f{q\pi (\hat{X}_1-\hat{X}_2)}{L}\right]}}{q\pi \epsilon}\right]}~&~\text{For}~ \beta \gg L\\
       \f{c}{3}\log{\left(\f{\beta}{\pi\epsilon}\right)}+\f{c L}{6 q \beta}\cdot\left[\f{\sin{\left[\f{q\pi (\hat{X}_1-\hat{X}_2)}{L}\right]}}{\sin{\left[\f{q\pi \hat{X}_1}{L}\right]}\sin{\left[\f{q\pi \hat{X}_2}{L}\right]}}\right] ~&~\text{For}~ L \gg \beta\\
   \end{cases},\\
    S_{A_2} &\approx   \f{c}{3}\log{\left(\f{\beta}{\pi\epsilon}\right)}+\f{c \pi Le^{2\theta}}{6 \beta}\cdot\f{l}{q}.
\end{split}
\ee
We kept the $\epsilon$-dependent terms because the two terms can compete in the last two expressions.

At the leading order of the large $\theta$ expansion in the high effective temperature region, for $\beta \gg L$, $S_{A_1}$ reduces to the vacuum entanglement entropy in the system with $L/q$, while for $L \gg \beta$, $S_{A_1}$ is proportional to $1/\beta$. Contrary to the entanglement entropy for the high-temperature thermal state on the flat spacetime, $S_{A_{1}}$ with  $L \gg \beta$ is not proportional to the subsystem size.
When the subsystem includes the points where the curvature is minimized, at the leading order in the large $\theta$ expansion, $S_{A_2}$ is proportional to the number of these points, and exponentially grows with $\theta$, not linearly. 
\subsubsection*{High temperature limit, $L/\beta \gg 1$}
Above, we considered the behavior of the entanglement entropy in the large $\theta$ region. 
Subsequently, we will explore the behavior of the entanglement entropy in the high temperature limit, $L/\beta \gg 1$, and consider the {$\theta$ dependence}.
In this limit, the theory-dependent piece of the entanglement entropy \eqref{eq:theEE} can be approximated as 
\be
\begin{split}
    S_{A_i}^{\text{T.D.}}&\approx  \f{c}{3}\log{\left(\f{\beta}{2\pi \epsilon}\right)}+\f{c L}{3\beta}   \cosh(2\theta)\cdot \f{ \left(\varphi_1-\varphi_2\right)}{q}.
\end{split}
\ee
We note that to ensure this expansion, we need the condition
\be\label{eq:condiExpAppro}
\begin{split}
   \f{L}{q\beta}   \cosh(2\theta) \left(\varphi_1-\varphi_2\right) \gg 1.
\end{split}
\ee

The dependence of the angle on the spatial locations, $X_{1}$ and $X_{2}$, is not clear in this expression.
To clarify this dependence, we focus on two limiting cases: $\theta \gg 1$; and $\theta \ll 1 $.
In the {large $\theta$} limit, $\theta \gg 1$, \eqref{eq:angleSin} results in the following approximation,
\be
\varphi_1-\varphi_2 \approx
\begin{dcases}
	e^{-2\theta} \cdot  \frac{\sin \left[\frac{q \pi\left(\hat{X}_1-\hat{X}_2\right)}{L}\right]}{\sin \left[\frac{q \pi \hat{X}_1}{L}\right] \sin \left[\frac{q \pi \hat{X}_2}{L}\right]} & m_{1}-m_{2}=0\\
	l \pi   & m_{1}-m_{2}=l\\
\end{dcases} \qquad \text{ for } \theta \gg 1.
\ee
Here, we note that to ensure this approximation, we can not consider cases where $X_{1}\sim L/q $ and/or  $X_{2}\sim 0 $. This can be seen by the expansion \eqref{eq:compPhiExpan}, and the sub-leading term (and other sub-sub-leading terms) can be order one in such cases. The extent to which $X_{1}$ and $X_{2}$ can approach $L/q$ and $0$, respectively, without the breakdown of the approximation, depends on the value of $\theta$. More concretely, here we need to impose the condition
\be \label{eq:condiAngleExpansion}
e^{-2\theta} \cdot  \frac{\sin \left[\frac{q \pi\left(\hat{X}_1-\hat{X}_2\right)}{L}\right]}{\sin \left[\frac{q \pi \hat{X}_1}{L}\right] \sin \left[\frac{q \pi \hat{X}_2}{L}\right]} \ll 1.
\ee
In the {small $\theta$} limit, $\theta \ll 1$,  the relation \eqref{eq:angle} implies 
\be
\varphi_1-\varphi_2 \approx \frac{q \pi (X_{1} - X_2)}{L}  \qquad \text{ for } \theta \ll 1.
\ee

\begin{figure}[th]
\begin{center}
\begin{tabular}{cc}
\subfigure[$L/\beta=10^{2}$]{
\includegraphics[scale=0.6]{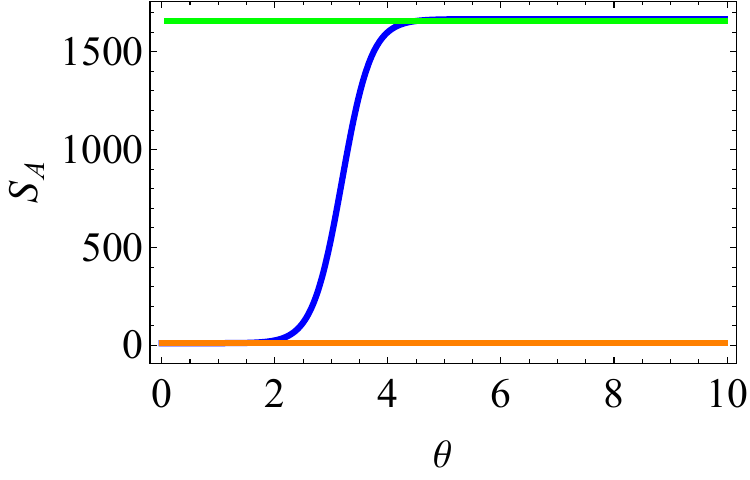}
} &
\subfigure[$L/\beta=10^{4}$]{
\includegraphics[scale=0.6]{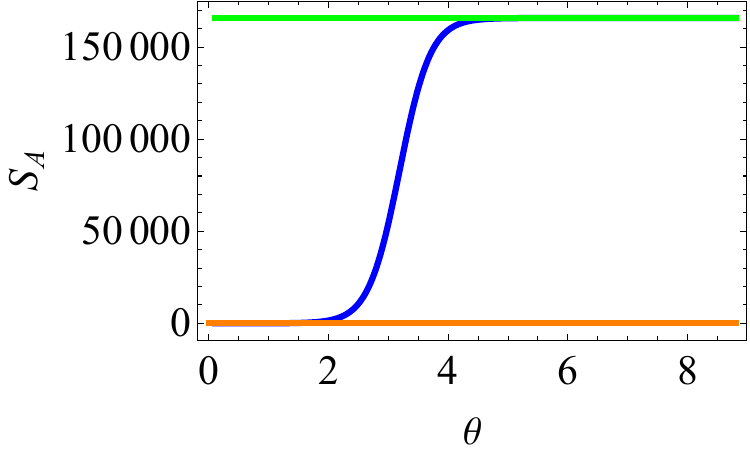}
}
\end{tabular}
\caption{Plots of the entropy as a function of $\theta$ for the region $A_{1}$. Blue curves correspond to the entanglement entropy for high temperature without approximation, orange curves do to the analytic expression for small $\theta$, i.e., the upper line of (\ref{eq:highLimitA1}), and green curves for large $\theta$, i.e., the lower line of (\ref{eq:highLimitA1}). Parameters: $q=4,L=100000,X_{1}=20,X_{2}=10, c=1, {\epsilon=10^{-14}}$. } 
\label{fig:entropyHighA1}
\end{center}
\end{figure}

\begin{figure}[ht]
\begin{center}
\begin{tabular}{cc}
\subfigure[$L/\beta=10^{2}$]{
\includegraphics[scale=0.6]{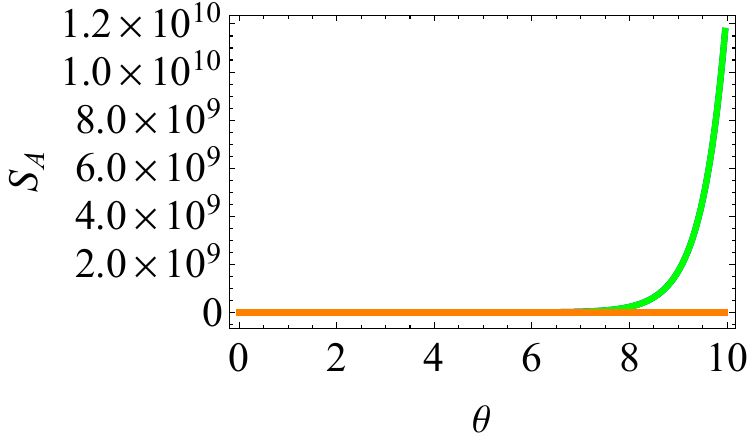}
} &
\subfigure[$L/\beta=10^{4}$]{
\includegraphics[scale=0.6]{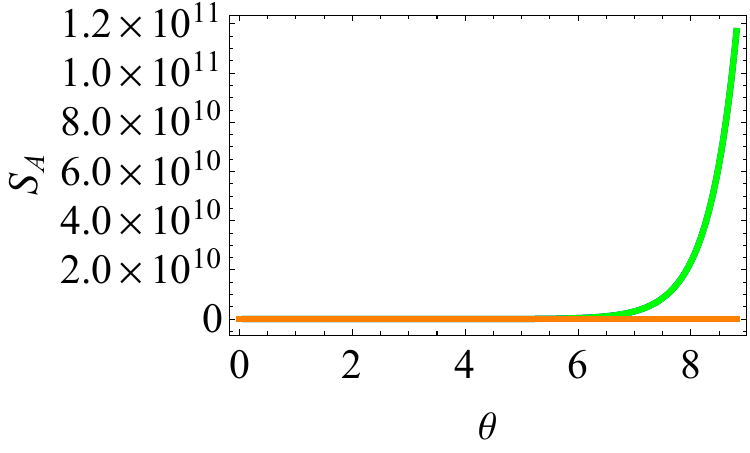}
}
\end{tabular}
\caption{{Plots of the entropy as a function of $\theta$ for the region $A_{2}$. Blue curves correspond to the entanglement entropy for high temperature without approximation, orange curves do to the analytic expression for small $\theta$, i.e., the upper line of (\ref{eq:highLimitA2}), and green curves for large $\theta$, i.e., the lower line of (\ref{eq:highLimitA2}). Parameters: $q=4,L=100000,X_{1}=2L/q+10,X_{2}=10,c=1, {\epsilon=10^{-14}}$.}} 
\label{fig:entropyHighA2}
\end{center}
\end{figure}

Combining these approximations, we obtain
\be\label{eq:highLimitA1}
\begin{aligned}
	S_{A_1}^{\text{T.D.}}&\approx
\begin{dcases}
\f{c}{3}\log{\left(\f{\beta}{2\pi\epsilon}\right)}+ \f{c \pi }{3\beta}  (\hat{X}_{1}-\hat{X}_{2}) & \theta \ll 1,\\
	\f{c}{3}\log{\left(\f{\beta}{2\pi\epsilon}\right)}+ \f{c L}{6q\beta} \cdot   \frac{\sin \left[\frac{q \pi\left(\hat{X}_1-\hat{X}_2\right)}{L}\right]}{\sin \left[\frac{q \pi \hat{X}_1}{L}\right] \sin \left[\frac{q \pi \hat{X}_2}{L}\right]} &  \theta \gg 1
\end{dcases}
\end{aligned}
\ee
and 
\be\label{eq:highLimitA2}
\begin{aligned}
	S_{A_2}^{\text{T.D.}}&\approx
\begin{dcases}
\f{c}{3}\log{\left(\f{\beta}{2\pi\epsilon}\right)}+ \f{c \pi }{3\beta} (X_{1}-X_{2}) 
 & \theta \ll 1,\\
	\f{c}{3}\log{\left(\f{\beta}{2\pi\epsilon}\right)}+ \f{c \pi  }{6\beta} \cdot \f{l}{q} L e^{2\theta}   &  \theta \gg 1
\end{dcases}
\end{aligned}
\ee
Of course, we need to consider contributions from the universal part, but they do not give significant contributions compared with the above theory-dependent parts.
In Figs. \ref{fig:entropyHighA1} and \ref{fig:entropyHighA2}, we show the dependence of the entanglement entropies on the general $\theta$.

\subsection{Entropy gap between two phase}\label{sec:entgap}

As explained in Section \ref{subsec:thermal-entropy}, the thermal entropy is discontinuous as the function of $\theta$ at the critical point \eqref{eq:critical-theta}, implying the first-order phase transition. In the following, we will investigate whether such discontinuities also exist for the entanglement entropies.

\begin{figure}[ht]
\begin{center}
\includegraphics[scale=0.8]{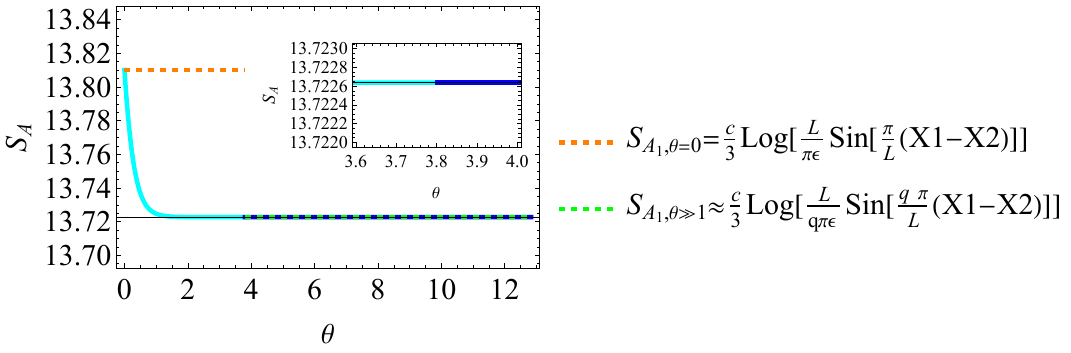}
\caption{{Plots of the entropy as a function of $\theta$ for the region $A_{1}$. Blue curves correspond to the expression for low temperature, and cyan curves do to that for high temperature. Parameters: $q=4,L=100000,L/\beta=10^{-3},X_{1}=17500,X_{2}=7500, c=1, {\epsilon=10^{-14}}$. Orange dashed curves correspond to the asymptotic behaviour for the limit $\theta \to 0$, and green dashed curves do to that for the limit {$\theta \gg 1$}. }}  
\label{fig:entropygapNumeA1}
\end{center}
\end{figure}

\begin{figure}[ht]
\begin{center}
\begin{tabular}{cc}
\subfigure[$L/\beta=10^{-2}$]{
\includegraphics[scale=0.6]{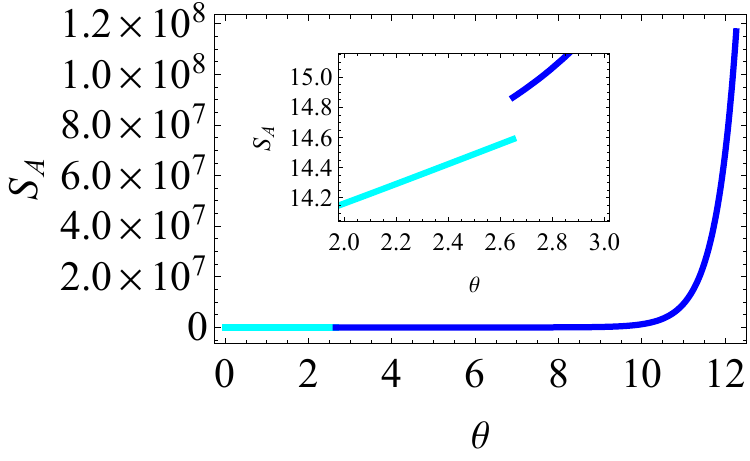}
} &
\subfigure[$L/\beta=10^{-5}$]{
\includegraphics[scale=0.6]{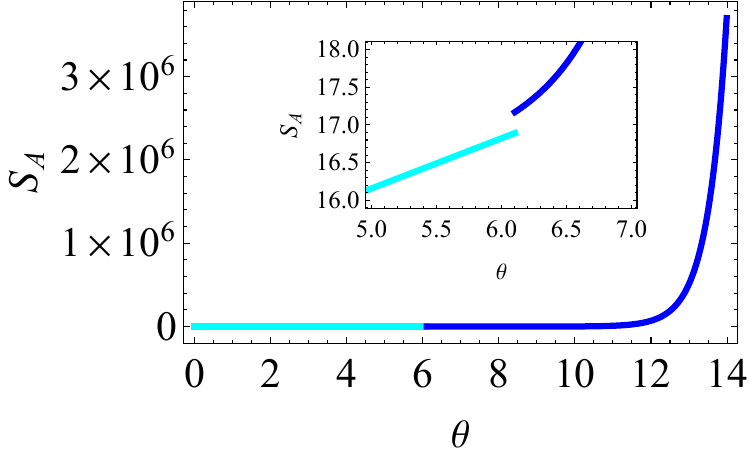}
}
\end{tabular}
\caption{Plots of the entanglement entropy as a function of $\theta$ for the region $A_{2}$. Blue curves illustrate the $\theta$ dependence of $S_{A_2}$ for low temperature, and cyan curves do for high temperature. Parameters considered here are $q=4,L=100000,X_{1}=2L/q+10^{3}+10,X_{2}=10^{3}+10,c=1, {\epsilon=10^{-14}}$.  }
\label{fig:entropygapNumeA2}
\end{center}
\end{figure}

First, under the assumption $L/\beta<1$, by plotting two entropies for low and high temperatures as a function of $\theta$ with fixing $L/\beta$, in Figs. \ref{fig:entropygapNumeA1} and \ref{fig:entropygapNumeA2}, we show the entanglement entropy as the function of $\theta$.

From Fig. \ref{fig:entropygapNumeA1} showing the {$\theta$ dependence} of  $S_{A_{1}}$, we can see that their two entropies are smoothly connected at the critical point \eqref{eq:critical-theta}
\footnote{Note that there is no critical point unless $L/\beta<1$.}. On the other hand, from Fig. \ref{fig:entropygapNumeA2} showing the $\theta$ dependence of $S_{A_{2}}$, we can see that there is a gap at the critical point. 

We define the entropy gap at the critical point as
\be\label{eq:defEntropyGap}
\begin{split}
	\Delta S_{A_{i}} &=\lim_{\theta\to \theta_{c}+0}S_{A_{i}}|_{\text{High}}-\lim_{\theta\to \theta_{c}-0}S_{A_{i}}|_{\text{Low}}\\
	&=\lim_{\theta\to \theta_{c}+0} S^{\text{T.D.}}_{A_{i}}|_{\text{High}}- \lim_{\theta\to \theta_{c}-0} S^{\text{T.D.}}_{A_{i}}|_{\text{Low}},
\end{split}
\ee
where in the second line, the universal terms cancel out.
Generally, it would be difficult to obtain an analytic expression of the gap at the critical point $\theta_{c}$. 
 However, by taking suitable limits, we can evaluate them analytically. One of such limits is $L/\beta \ll 1$ limit. In this case, the critical point $\theta_{c}$ is approximately given by \eqref{eq:critical-theta},
 \begin{equation*}
     \theta_{c} \approx \frac{1}{2}\log\left( \frac{2\beta}{L} \right)\gg 1,
 \end{equation*}
 which is very large.
 This is the same as the critical point for the thermal entropy in the $L/\beta \ll 1$.
 Thus, since $\theta_c$ is much larger than one, we will evaluate the entropy gap in the large $\theta$ expansion.
 At the leading order in the large $\theta$ expansion, for the region $A_{1}$, there are no entropy gaps,
\be\label{eq:EEGapA1Limit}
\begin{split}
	\Delta S_{A_{1}} &=S^{\text{T.D.}}_{A_{1}}|_{\text{High}}-S^{\text{T.D.}}_{A_{1}}|_{\text{Low}}\\
	&\approx  0.
\end{split}
\ee

On the other hand, at the leading order in the large $\theta$ expansion, for the region $A_{2}$, there is a non-zero entropy gap. Let us find the analytic expression of the entropy gap. 
We can not use the expression \eqref{eq:EEforHigh-LargeTheta} for $S_{A_{2}}$ since we take the limit, where $L/\beta \ll 1$, and $\theta \gg 1$, with fixing $L \cosh (2\theta)/\beta=1$.  In this limit, the entanglement entropy $S_{A_{2}}$, \eqref{eq:EEforHigh-LargeTheta}, should be approximated as 
\begin{equation}
    S_{A_{2}}  \approx \f{c}{3}\log{\left(\f{\beta}{\pi \epsilon}\right)} +  \text{Min}\left[ \f{c}{3}\log{ \sinh\left( \f{l}{q}\pi\right)  } , \f{c}{3}\log{ \left[ \sinh\left( \left( 1\mp \f{l}{q} \right)\pi\right) \right] } + \f{c\pi}{3}  \right]
\end{equation}
at the leading order in the large $\theta$ expansion.

\begin{figure}[ht]
\begin{center}
\begin{tabular}{cc}
\subfigure[$q=4$]{
\includegraphics[scale=0.6]{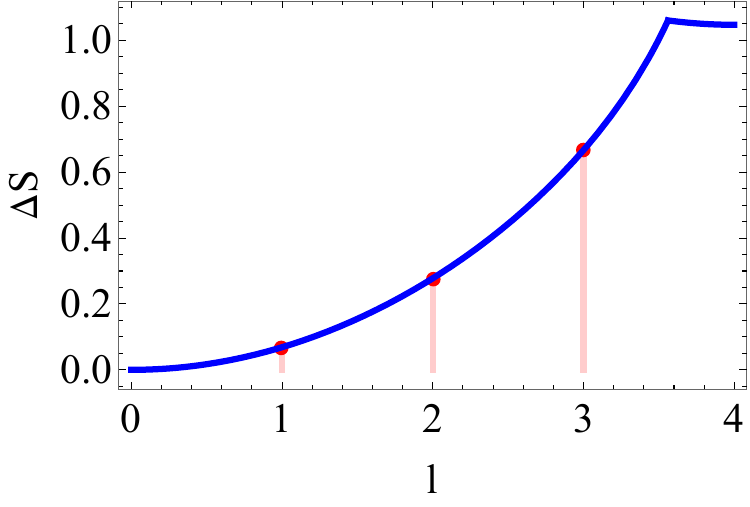}
} &
\subfigure[$q=200$]{
\includegraphics[scale=0.6]{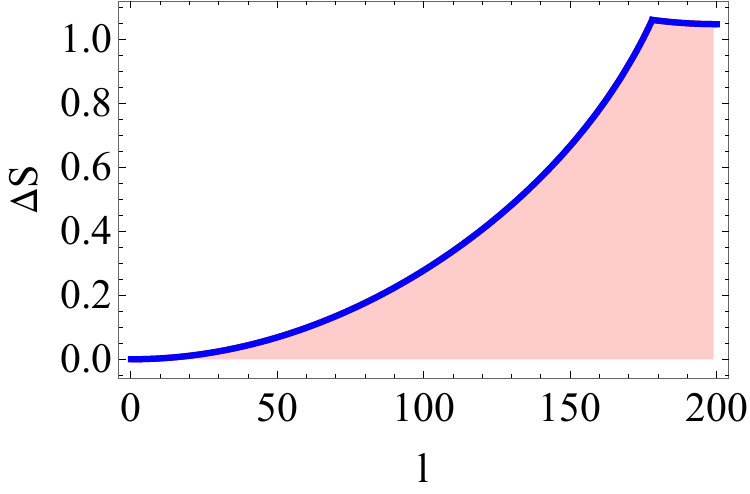}
}
\end{tabular}
\caption{Plots of the entropy gap and its analytic expression (\ref{eq:entropygap}) as a function of $l$, which is related to $X_{1}$ by $X_{1}=L\cdot l/q+X_{2}$. Blue solid lines are given by the analytic expression, and red dots are entropy gap obtained by using the definitions  (\ref{eq:defEntropyGap}), (\ref{eq:EEvac}), (\ref{eq:theEE}) and (\ref{eq:angle}) by numerical evaluations of the gap without using the large $\theta$ approximation. Parameters are chosen to be $L=100000,X_{2}=10,\beta =1000L,c=1$. }
\label{fig:entropygapCompa}
\end{center}
\end{figure}

 Using this expression with \eqref{eq:nonUniEELow}, we obtain the entropy gap under the limit,
 \be
\begin{split}\label{eq:entropygap}
	\Delta S_{A_{2}} & \approx \left[\f{c}{3}\log{\left(\f{\beta}{\pi\epsilon}\right)} +  \text{Min}\left\{ \f{c}{3}\log{ \sinh\left( \f{l}{q}\pi\right)  } , \f{c}{3}\log{ \sinh\left( \left( 1\mp \f{l}{q} \right)\pi\right)  } + \f{c\pi}{3}  \right\}  \right] \\ 
	& \hspace{5cm}  -  \left[ \f{c}{3}\log{\left(\f{L e^{2\theta_{c}}}{2\pi\epsilon}\right)}+\f{c}{3}\log{ \sin\left( \f{l}{q}\pi\right)  } \right]\\
	&= \f{c}{3}  \text{Min} \left\{ \log{ \left[\f{ \sinh\left( \dfrac{l}{q}\pi\right) }{\sin\left( \dfrac{l}{q}\pi\right)}  \right]} , \log{\left[ \f{ \sinh\left( \left(1- \dfrac{l}{q} \right)\pi\right) }{\sin\left( \dfrac{l}{q}\pi\right)}  \right]} + \pi  \right\} .
\end{split}
\ee
This analytic result under the limit matches the entropy gap shown in Fig. \ref{fig:entropygapCompa}.


We give some comments on the above result. First, the above minimum value is given by the first factor for smaller $l$, and the second factor for larger $l\, (\leq q-1)$.

\begin{figure}[ht]
\begin{center}
\includegraphics[scale=0.6]{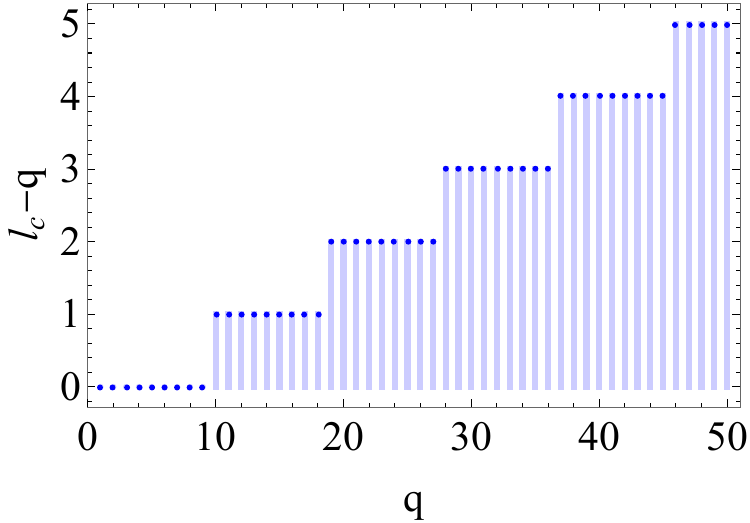}
\caption{Plot of the critical value $l_{c}$ (\ref{eq:critil}) minus $q$, $l_{c}-q$,  as a function of $q$. }
\label{fig:critilPlot}
\end{center}
\end{figure}

More precisely, the gap is given by 
\begin{equation}
    \Delta S_{A_{2}}  \approx \begin{dcases}
        \f{c}{3} \log{ \left[\f{ \sinh\left( \dfrac{l}{q}\pi\right) }{\sin\left( \dfrac{l}{q}\pi\right)}  \right]},  & \quad \text{ for }  1\leq l \leq  l_{c}-1\\
        \f{c}{3} \log{\left[ \f{ \sinh\left( \left(1- \dfrac{l}{q} \right)\pi\right) }{\sin\left( \dfrac{l}{q}\pi\right)}  \right]} + \f{c\pi}{3},  &  \quad \text{ for } l_{c}-1 < l \leq   q-1,
    \end{dcases}
\end{equation}
where $l_{c}$ is  the critical value given by
\begin{equation}
	l_{c}=  \left\lceil  \f{q}{2\pi} \log \left( \f{e^{2\pi}+1}{2}  \right) \right\rceil.\label{eq:critil}
\end{equation}
Here, $\lceil x \rceil$ is the ceiling function, and $\f{1}{2\pi} \log \left( \f{e^{2\pi}+1}{2}  \right) = 0.889979 \cdots  $.
By plotting $l_c$ as a function of $q$ (see Fig. \ref{fig:critilPlot}), we can see the following behavior of $l_c$,
\begin{equation}
	l_{c}=\begin{cases}
		 q &  1 \leq q \leq 9,\\
		 q-1 & 10 \leq  q \leq 18,\\
		 q-2 & 19 \leq  q \leq 27,\\
		 q-3 & 28 \leq  q \leq 36,\\
		 \cdots.
	\end{cases}
\end{equation}
The $q$ dependence of $l_c$ can be divided into twofold: for $q\le 9$, $l_c=q$; and for $q>9$, $q>l_c$. 
Therefore, for the former case, the entropy gap is determined by the first term in (\ref{eq:entropygap}). For the latter case with $l<l_c$, the gap is determined by the first term, while for that with $l_c<l$, it is determined by the second term. 
We note that, for smaller $q$, the above behavior is simply related to multiples of $9$ due to $\f{1}{2\pi} \log \left( \f{e^{2\pi}+1}{2}  \right) = 0.889979 \cdots \approx 0.9 $, but for larger $q$, there is a deviation between
the exact value of $l_{c}$ evaluated by the definition
and the estimation based on smaller $q$. 

In this result, by taking the limit $l\to 0$ (with analytic continuation of integers $l$ to real numbers), which implies that the region $A_{2}$ reduces to the region $A_{1}$, clearly $\Delta S_{A_{2}}$ vanishes,
 \be
\begin{split}
	\lim_{l\to 0}  \Delta S_{A_{2}}  &= \lim_{l\to 0} \f{c}{3} \left[ \log{ \f{ \sinh\left( \dfrac{l}{q}\pi\right) }{\sin\left( \dfrac{l}{q}\pi\right)}  }\right]\\
	&= 0.
\end{split}
\ee
Relatedly, if we take the limit $q\to \infty$, then for $l =\mathcal{O}(q^{0})$, we obtain the vanishing gap again,
 \be
\begin{split}
	\lim_{q\to \infty}  \Delta S_{A_{2}}  &= \lim_{q\to \infty} \f{c}{3} \left[ \log{ \f{ \sinh\left( \dfrac{l}{q}\pi\right) }{\sin\left( \dfrac{l}{q}\pi\right)}  }\right]\\
	&= 0.
\end{split} \qquad \text{ for } l =\mathcal{O}(q^{0}),
\ee

\subsection{Mutual information \label{eq:mutual-informaion}}

We compute mutual information to study the $\theta$ dependence of non-local correlation.
We divide the $\theta$ dependence of the mutual information into the two regions, the high and low effective temperature region, and consider the dependence of the mutual information.
 In the low effective temperature regime, $\tau_{\text{Mod.}}<1$, at the leading order in the large $\theta$ expansion, as obtained in \eqref{eq:EE-Low-LargeTheta}, the entanglement entropy for the single interval in $2$d holographic CFTs reduces to the vacuum one when the interval does not contain the fixed points.
 When the interval contains the fixed points the single-interval entanglement entropy becomes the constant value, which is independent of the interval size.
 Therefore, when both of the intervals do not include the fixed points, the entanglement entropy even for the double interval is expected to reduce to the vacuum one in the large $\theta$ limit. 
 This suggests that the mutual information of two nearby intervals can be non-zero, but that of the two distant regions would be zero.
 When both of the intervals include the fixed points, the entanglement entropy for the double interval is given by twice the constant value for the single interval, leading to vanishing mutual information.
On the other hand, in the high effective temperature regime, $\tau_{\text{Mod.}}>1$, as in \eqref{eq:high-temperature-region-in-large-theta}, the entanglement entropy for holographic CFT is given by thermal entanglement entropy at the leading order in the large $\theta$ expansion, when the intervals contain the fixed points. Naively, this would give a vanishing mutual information. However, when the intervals do not include the fixed points, the behavior of the entanglement entropy changes depending on whether $L/\beta \ll 1$ or $L/\beta \gg 1$ as in \eqref{eq:high-temperature-region-in-large-theta}: it reduces to the vacuum one in the former case, and to the thermal-like one in the latter case. Thus, the resulting mutual information depends on these limits, and naively it would be given by the vacuum mutual information in $L/\beta \ll 1$ and the vanishing one in $L/\beta \gg 1$. 
  Let us study the details of the mutual information in these various situations.
  
We explore the properties of the mutual information between the subsystems that do not contain the points where the absolute value of the curvature is maximized.
To do so, we define the new subsystem $A'$ as
 \be
A'=\left\{x| 0<X_3<x<X_4\right\},
\ee
where we assume that
\be
   \f{(n+1)L}{q}>X_3>X_4>\f{nL}{q}~~\text{For}~A_1'.
\ee
Then, we explore the $\theta$ and temperature dependence of the mutual information between $A_1$ and $A'_1$, $I(A_1,A'_1)$.
The subsystems, $A_1$ and $A'_1$, are in the spatial interval between $(n+1)L/q$ and $nL/q$.
The mutual information, $I(A_1,A'_1)$ , is given by
\begin{equation}\label{eq:mutualA1A1'}
	I(A_{1},A_{1}') = S_{A_1}^{\text{T.D.}} + S_{A_1'}^{\text{T.D.}} - S_{A_1 \cup A_1'}^{\text{T.D.}},
\end{equation}
where the universal parts are canceled out. 
Since the first and second terms of (\ref{eq:mutualA1A1'}) are already obtained in the previous discussions, we focus on the theory-dependent piece, $S_{A_1 \cup A_1'}^{\text{T.D.}}$. In 2d holographic CFT, the theory-dependent piece is determined by
\begin{equation}
	S_{A_1 \cup A_1'}^{\text{T.D.}} \approx \text{Min}\left[ S_{A_1}^{\text{T.D.}} + S_{A_1'}^{\text{T.D.}}, S_{A_1\cup B_{A_1;A_1} \cup A_1'}^{\text{T.D.}} + S_{B_{A_1;A_1}}^{\text{T.D.}} \right],
\end{equation}
where $B_{A_1;A_1}$ denotes the subsystems defined by
\begin{equation}
	B_{A_1;A_1} =\left\{x| 0<X_1<x<X_4\right\}.
\end{equation}
Since each of these entanglement entropies can be computed by using the previous discussions, we can evaluate this mutual information. 

\begin{figure}[ht]
\begin{center}
\includegraphics[scale=0.6]{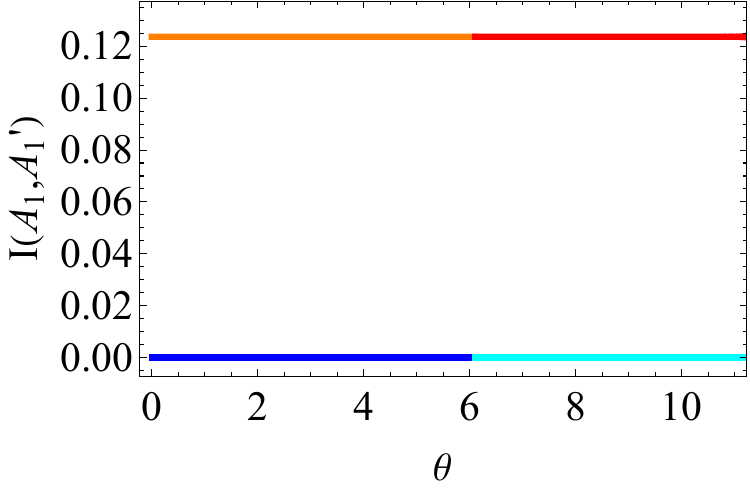}
\caption{Plot of the mutual information (\ref{eq:mutualA1A1'}) as a function of $\theta$ in the low temperature region, $L/\beta<1$. Blue curve illustrates the $\theta$ dependence of $I(A_1,A'_1)$ for the two distant subsystems in the low effective temperature region, and cyan curve illustrates that in the high effective temperature. For these curves, parameters are set to be $q=4,L=100000,{L/\beta=10^{-5}},X_{1}=10^{3}+20,X_{2}=10^{3}+10,X_{3}=10^{4}+X_{1},X_{4}=10^{4}+X_{2}, c=1 $. Orange curve illustrates the $\theta$ dependence of $I(A_1,A'_1)$ for the two nearby subsystems in the low effective temperature, and red curve illustrates that in the high effective temperature region. For these curves, parameters are set to be  $q=4,L=100000,L/\beta=10^{-5},X_{1}=10^{3}+20,X_{2}=10^{3}+10,X_{3}=X_{1}+13,X_{4}=X_{2}+13, c=1 $. 
}
\label{fig:MutualA1}
\end{center}
\end{figure}

\begin{figure}[ht]
\begin{center}
\begin{tabular}{cc}
\subfigure[Subsystems near unstable fixed points]{
\includegraphics[scale=0.6]{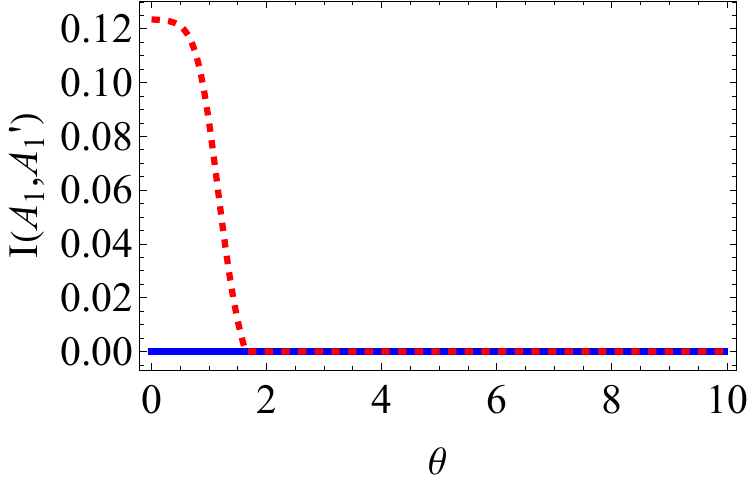}
} &
\subfigure[Subsystems near stable fixed points]{
\includegraphics[scale=0.6]{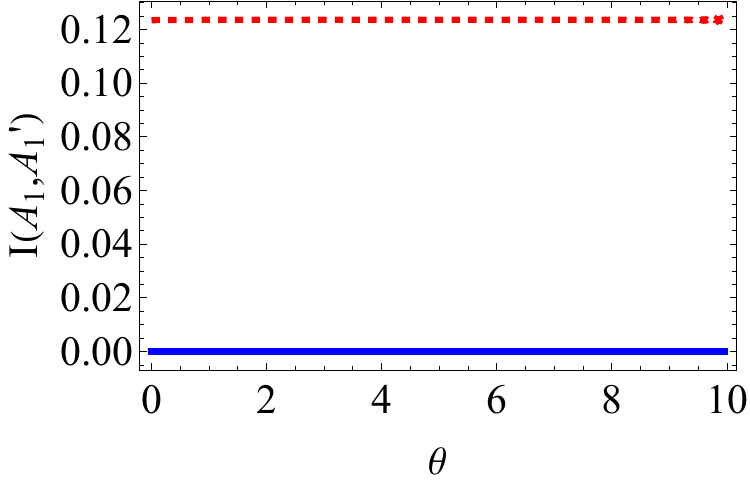}
}
\end{tabular}
\caption{Plots of the mutual information (\ref{eq:mutualA1A1'}) as a function of $\theta$ in the high temperature region, $L/\beta>1$. Blue curves illustrate the $\theta$ dependence of $I(A_1,A_1')$ for two distant subsystems. For these curves, parameters are set to be $q=4,L=100000,L/\beta=10^{2},c=1$, and for the left panel $X_{1}=10^{3}+20,X_{2}=10^{3}+10,X_{3}=10^{4}+X_{1},X_{4}=10^{4}+X_{2}$ and for the right panel $X_{1}=12400+10,X_{2}=12400,X_{3}=12400+10^{4}+10,X_{4}=12400+10^{4}$. Red dashed curves illustrate the $\theta$ dependence of $I(A_1,A_1')$ for two nearby subsystems. For these curves, parameters are again set to be  $q=4,L=100000,L/\beta=10^{2}, c=1$, and for the left panel, $X_{1}=10^{3}+20,X_{2}=10^{3}+10,X_{3}=X_{1}+13,X_{4}=X_{2}+13 $ and for the right panel, $X_{1}=12400+10,X_{2}=12400,X_{3}=12400+3+20,X_{4}=12400+3+10$.} 
\label{fig:MutualA1High}
\end{center}
\end{figure}

In Fig. \ref{fig:MutualA1}, we show the plots of the mutual information (\ref{eq:mutualA1A1'}) as a function of $\theta$ in the low temperature region, $L/\beta<1$. 
In the plots, the subsystems for which mutual information is computed are two-fold: two distant subsystems; and two nearby ones. 
In Fig. \ref{fig:MutualA1High}, we show the plots of the mutual information (\ref{eq:mutualA1A1'}) as a function of $\theta$ in the high temperature region, $L/\beta>1$.
The subsystems considered in these figures are the same as those in Fig. \ref{fig:MutualA1}.

In general, it is difficult to obtain analytic expressions of correlation length where the mutual information vanishes. However, by taking suitable limits, we can obtain their analytic expression. 
To understand how their correlation lengths depend on $\theta$, let us focus on several limiting cases below.

\subsubsection{Low effective temperature regime \texorpdfstring{$L_\text{eff}/\beta \leq 1$}{Leff/beta<=1}}
First, we focus on $I(A_1,A_1')$ in the low temperature regimes where $\tau_\text{Mod.}=L_\text{eff}/\beta \leq 1$. In this regime, the entanglement entropies for the intervals considered here reduce to the one, resembling the vacuum entanglement entropy, as in \eqref{eq:EEvac}.
Then, the mutual information \eqref{eq:mutualA1A1'} can be written as
\begin{equation}\label{eq:MILowGeneral}
	I(A_{1},A_{1}') =  \max\left\{ 0, \frac{c}{3} \log\left[ \frac{ \sin \left( \frac{\varphi_{1}-\varphi_{2}}{q} \right) \sin \left( \frac{\varphi_{3}-\varphi_{4}}{q} \right) }{ \sin \left( \frac{\varphi_{3}-\varphi_{2}}{q} \right) \sin \left( \frac{\varphi_{4}-\varphi_{1}}{q} \right) } \right] \right\}.
\end{equation}
Thus, since the behavior of $I(A_{1},A_{1}')$ is determined by the ratio in the logarithm, below in several limiting cases, we closely look at this ratio,
\begin{equation}\label{eq:ratioLow}
	\frac{ \sin \left( \frac{\varphi_{1}-\varphi_{2}}{q} \right) \sin \left( \frac{\varphi_{3}-\varphi_{4}}{q} \right) }{ \sin \left( \frac{\varphi_{3}-\varphi_{2}}{q} \right) \sin \left( \frac{\varphi_{4}-\varphi_{1}}{q} \right) }.
\end{equation}

\subsubsection*{Low temperature \texorpdfstring{$L/\beta<1$}{L/beta<1}  and \texorpdfstring{$\theta\to 0$}{theta->0} limit}
We focus on the ratio \eqref{eq:ratioLow} in the parameter region, $L/\beta<1$ and $\theta\to  0$. 
We note that this parameter region can include the region, where $L/\beta \ll 1$ and $\theta\to  0$.
In this case, by using \eqref{eq:angle}, we can approximate the angle function $\varphi$ by
\begin{equation}\label{eq:angleSmalltheta}
	\varphi_{i} = \frac{\pi q}{L}X_{i}  \quad \text{for } \theta \to 0. 
\end{equation}
Using this approximation, we obtain
\begin{equation}
	\begin{aligned}
	&\sin \left( \frac{\varphi_{i}-\varphi_{j}}{q} \right)=  \sin\left( \frac{\pi}{L}(X_{i}-X_{j}) \right) .
	\end{aligned}
\end{equation}
Then, the ratio \eqref{eq:ratioLow} becomes
\begin{equation}\label{eq:ratioLowExpand}
	\frac{ \sin \left( \frac{\varphi_{1}-\varphi_{2}}{q} \right) \sin \left( \frac{\varphi_{3}-\varphi_{4}}{q} \right) }{ \sin \left( \frac{\varphi_{3}-\varphi_{2}}{q} \right) \sin \left( \frac{\varphi_{4}-\varphi_{1}}{q} \right) } = \frac{ \sin\left( \frac{\pi}{L}(X_{1}-X_{2}) \right) \sin\left( \frac{\pi}{L}(X_{3}-X_{4}) \right) }{ \sin\left( \frac{\pi}{L}(X_{3}-X_{2}) \right) \sin\left( \frac{\pi}{L}(X_{4}-X_{1}) \right) },
\end{equation}
and the mutual information reduces to the vacuum one for the system with the size of $L$,
\begin{equation}
	I(A_{1},A_{1}') =  \max\left\{ 0, \frac{c}{3} \log\left[ \frac{ \sin\left( \frac{\pi}{L}(X_{1}-X_{2}) \right) \sin\left( \frac{\pi}{L}(X_{3}-X_{4}) \right) }{ \sin\left( \frac{\pi}{L}(X_{3}-X_{2}) \right) \sin\left( \frac{\pi}{L}(X_{4}-X_{1}) \right) } \right] \right\}.
\end{equation}


\subsubsection{High temperature regime \texorpdfstring{$L_\text{eff}/\beta \geq 1$}{Leff/beta>=1}}

Then, we move on the study of the mutual information for high effective temperature regimes $\tau_\text{Mod.}=L_\text{eff}/\beta \geq 1$.
In these regimes, by taking suitable limits, we can analytically explore $I(A_1,A_1')$. 
In the high temperature regimes, the entanglement entropy is given by \eqref{eq:theEE}. Using this expression, the mutual information can be given in the following simple form\footnote{Since we focus on the small intervals $A_{1},A_{1}'$ compared to the total system, we simply ignore some pieces of contributions appearing in \eqref{eq:theEEOrigi}, which would give smaller contributions than those appearing here.},
\begin{equation}\label{eq:mutualHigh}
	I(A_{1},A_{1}') =  \max\left\{ 0, \frac{c}{3} \log\left[ \frac{ \sinh{\left[\f{L_{\text{eff}}\left(\varphi_1-\varphi_2\right)}{q\beta}\right]}  \sinh{\left[\f{L_{\text{eff}}\left(\varphi_3-\varphi_4\right)}{q\beta}\right]} }{ \sinh{\left[\f{L_{\text{eff}}\left(\varphi_3-\varphi_2\right)}{q\beta}\right]} \sinh{\left[\f{L_{\text{eff}}\left(\varphi_4-\varphi_1\right)}{q\beta}\right]} } \right] \right\}.
\end{equation}
Thus, to study properties of this mutual information, we focus on the cross ratio
\begin{equation}\label{eq:ratioHigh}
	\frac{ \sinh{\left[\f{L_{\text{eff}}\left(\varphi_1-\varphi_2\right)}{q\beta}\right]}  \sinh{\left[\f{L_{\text{eff}}\left(\varphi_3-\varphi_4\right)}{q\beta}\right]} }{ \sinh{\left[\f{L_{\text{eff}}\left(\varphi_3-\varphi_2\right)}{q\beta}\right]} \sinh{\left[\f{L_{\text{eff}}\left(\varphi_4-\varphi_1\right)}{q\beta}\right]} },
\end{equation}
and below we evaluate it under several limits.

\subsubsection*{Low temperature and large \texorpdfstring{$\theta$}{theta} limit, \texorpdfstring{$L/\beta \ll 1$}{L/beta<<1} and $\theta \gg 1$,  with  \texorpdfstring{$L_\text{eff}/\beta \gtrsim 1$}{Leff/beta>=1}}

First, we evaluate the ratio \eqref{eq:ratioHigh} in parameter region, where $L/\beta \ll 1$, $\theta \gg 1$, and $L_\text{eff}/\beta \gtrsim 1$. 
In this parameter region, since the angle function $\varphi_{i}$ can be approximated by
 \eqref{eq:AngleExpanLargeTheta}, we can approximate a constitute piece of the ratio as
 \begin{equation}
 	\begin{aligned}
 		&\sinh{\left[\f{L_{\text{eff}}\left(\varphi_i-\varphi_j\right)}{q\beta}\right]} \approx \sinh \left( \frac{L}{2q\beta} \left[\cot\left(\f{q \pi X_{j}}{L}\right) - \cot\left(\f{q \pi X_{i}}{L}\right)   \right] \right). 	\end{aligned}
 \end{equation}
  We note that the approximation is valid only under the condition $e^{-2\theta}\cot\left(\f{q \pi \hat{X}_{i}}{L}\right)  \ll 1$, implying that we can not apply the above expression for intervals near unstable fixed points, $\hat{X}_{i} \sim 0, L/q$.
  Later, we study such cases. 
In the low temperature region considered here, we can further simplify the above expression as follows,
 \begin{equation}
 	\begin{aligned}
 		\sinh{\left[\f{L_{\text{eff}}\left(\varphi_i-\varphi_j\right)}{q\beta}\right]} &\approx  \frac{L}{2q\beta} \left[\cot\left(\f{q \pi X_{j}}{L}\right) - \cot\left(\f{q \pi X_{i}}{L}\right)   \right]\\
 		&=\frac{L}{2q\beta} \cdot \frac{ \sin\left(\f{q \pi }{L} (X_{i} -X_{j})\right)  }{\sin\left(\f{q \pi X_{i}}{L}\right) \sin\left(\f{q \pi X_{j}}{L}\right)}.
 	\end{aligned}
 \end{equation}
 Then, the ratio \eqref{eq:ratioHigh} becomes 

\begin{equation}\label{eq:approRatioHighSuperLow}
	\begin{aligned}
		&\frac{ \sinh{\left[\f{L_{\text{eff}}\left(\varphi_1-\varphi_2\right)}{q\beta}\right]}  \sinh{\left[\f{L_{\text{eff}}\left(\varphi_3-\varphi_4\right)}{q\beta}\right]} }{ \sinh{\left[\f{L_{\text{eff}}\left(\varphi_3-\varphi_2\right)}{q\beta}\right]} \sinh{\left[\f{L_{\text{eff}}\left(\varphi_4-\varphi_1\right)}{q\beta}\right]} } \approx \frac{ \sin\left( \frac{q\pi}{L}(X_{1}-X_{2}) \right) \sin\left( \frac{q\pi}{L}(X_{3}-X_{4}) \right) }{ \sin\left( \frac{q\pi}{L}(X_{3}-X_{2}) \right) \sin\left( \frac{q\pi}{L}(X_{4}-X_{1}) \right) }. 
	\end{aligned}
\end{equation}
We note that the previous result \eqref{eq:ratioLowExpand} indicates that the system is given by $L$, but the current result implies that the effective system size is given by $L/q$. This is consistent with the intuition that for small $\theta$, the system is almost given by a original single interval with size $L$, but for large $\theta$, the system is almost decoupled into $q$ smaller intervals with size $L/q$. 





\subsubsection{ High effective temperature limit \texorpdfstring{$L_\text{eff}/\beta \gg 1$}{Leff/beta >>1}}

We focus on the high effective temperature and explore the mutual information in (\ref{eq:mutualA1A1'}).
We start with the expression \eqref{eq:mutualHigh} and expand the ratio in \eqref{eq:mutualHigh} with respect to $\f{L}{\beta}\gg 1$, while keeping $q$ a constant not propotional to the ratio $\f{L}{\beta}$. 
At the leading order in this expansion, this ratio is approximately given by
\begin{equation}\label{eq:approRatioSuperHi}
	\begin{aligned}
		\frac{ \sinh{\left[\f{L_{\text{eff}}\left(\varphi_1-\varphi_2\right)}{q\beta}\right]}  \sinh{\left[\f{L_{\text{eff}}\left(\varphi_3-\varphi_4\right)}{q\beta}\right]} }{ \sinh{\left[\f{L_{\text{eff}}\left(\varphi_3-\varphi_2\right)}{q\beta}\right]} \sinh{\left[\f{L_{\text{eff}}\left(\varphi_4-\varphi_1\right)}{q\beta}\right]} } \approx \f{1}{\exp\left[\f{2L_{\text{eff}}\left(\varphi_4-\varphi_1\right)}{q\beta}\right]-1},
	\end{aligned}
\end{equation}
where we assumed that 
\begin{equation}\label{eq:approConditionSuperHigh}
	\f{L_{\text{eff}}\left(\varphi_1-\varphi_2\right)}{q\beta},\f{L_{\text{eff}}\left(\varphi_3-\varphi_4\right)}{q\beta},\f{L_{\text{eff}}\left(\varphi_3-\varphi_2\right)}{q\beta} \gg 1.
\end{equation}
Then, we evaluate the angle difference $\varphi_4-\varphi_1$ to give an analytic expression for the mutual information.
For convenience, let us discuss the $\theta \to 0$ case.
In the $\theta \to 0$ limit, the angle difference $\varphi_4-\varphi_1$ is simply given by
\begin{equation}
	\varphi_4-\varphi_1 \to  \f{\pi q}{L}\left( X_{4} -X_{1} \right),
\end{equation}
and the ratio becomes 
\begin{equation}
	\f{1}{\exp\left[\f{2L_{\text{eff}}\left(\varphi_4-\varphi_1\right)}{q\beta}\right]-1} \to  \f{1}{\exp\left[ \f{2\pi}{\beta}\left( X_{4} -X_{1} \right)  \right]-1}.
\end{equation}
Thus, the mutual information reduces to the usual result,
\begin{equation}
	\left.I(A_{1},A_{1}')\right|_{\text{High temperature limit w/ } \theta=0} =  \max\left\{ 0, - \frac{c}{3} \log\left[ \exp\left[ \f{2\pi}{\beta}\left( X_{4} -X_{1} \right)  \right]-1 \right] \right\}.
\end{equation}
From this mutual information, we can read off the thermal correlation length $d_{41}$ defined through the condition
\begin{equation}
	\frac{c}{3} \log\left[ \exp\left[ \f{2\pi}{\beta} d_{41}   \right]-1 \right] =0,
\end{equation}
resulting in the uniform thermal correlation length,
\begin{equation}
	d_{41,\theta=0}=\f{\beta}{2\pi}\log 2.
\end{equation}

\subsubsection*{High temperature, \texorpdfstring{$L/\beta\gg 1$}{L/beta>>1}, and large \texorpdfstring{$\theta$}{theta} limit}

In the large $\theta$ limit, we evaluate the angle difference $\varphi_4-\varphi_1$. At the leading order in the expansion in \eqref{eq:AngleExpanLargeTheta}, we have 
\begin{equation}
	\begin{aligned}
		\exp\left[\f{2L_{\text{eff}}\left(\varphi_1-\varphi_4\right)}{q\beta}\right] -1 &\approx  \exp\left[\f{L}{q\beta} \left(\cot\left( \f{q\pi}{L}\hat{X}_{4} \right)-\cot\left( \f{q\pi}{L}\hat{X}_{1} \right)\right)\right] -1\\
		& \approx  \exp\left[\f{L}{q\beta} \cdot \f{\sin\left( \f{q\pi}{L}(\hat{X}_{4}-\hat{X}_{1}) \right) }{\sin\left( \f{q\pi}{L}\hat{X}_{1} \right) \sin\left( \f{q\pi}{L}\hat{X}_{4} \right)} \right] -1\\
		&= \exp\left[\f{2L}{q\beta} \cdot \f{\sin\left( \f{q\pi}{L}(\hat{X}_{4}-\hat{X}_{1}) \right) }{\cos\left( \f{q\pi}{L}(\hat{X}_{4}-\hat{X}_{1}) \right) - \cos\left( \f{q\pi}{L}(\hat{X}_{1}+\hat{X}_{4}) \right)} \right] -1.
	\end{aligned}
\end{equation}
This gives the mutual information,
\begin{equation}
	\begin{aligned}
		&\left.I(A_{1},A_{1}')\right|_{\text{High temperature limit w/ } \theta\gg 1}\\
		& \approx  \max\left\{ 0, -\frac{c}{3} \log\left[\exp\left[\f{2L}{q\beta} \cdot \f{\sin\left( \f{q\pi}{L}(\hat{X}_{4}-\hat{X}_{1}) \right) }{\cos\left( \f{q\pi}{L}(\hat{X}_{4}-\hat{X}_{1}) \right) - \cos\left( \f{q\pi}{L}(\hat{X}_{1}+\hat{X}_{4}) \right)} \right] -1 \right] \right\}.
	\end{aligned}
\end{equation}
From the mutual information, we can define the induced correlation length $d_{41,\theta\gg 1}$ defined by the condition,
\begin{equation}
	-\frac{c}{3} \log\left[\exp\left[\f{2L}{q\beta} \cdot \f{\sin\left( \f{q\pi}{L}d_{41,\theta\gg 1} \right) }{\cos\left( \f{q\pi}{L}d_{41,\theta\gg 1}\right) - \cos\left( \f{q\pi}{L}(\hat{X}_{1}+\hat{X}_{4}) \right)} \right] -1 \right] =0.
\end{equation}
This condition can be written as
\begin{equation}
	\begin{aligned}
		\f{\sin\left( \f{q\pi}{L}d_{41,\theta\gg 1} \right) }{\cos\left( \f{q\pi}{L}d_{41,\theta\gg 1}\right) - 1+ f_{\theta\to \infty,X_{1,4}} } = \f{q\beta}{2L}\log 2,
	\end{aligned}
\end{equation}
where we used the definition of the envelop function \eqref{eq:enveFunc} and defined $f_{\theta\to \infty,X_{1,4}}$ as 
\begin{equation}
	f_{\theta\to \infty,X_{1,4}} \coloneqq f\left( \f{(\hat{X}_{1}+\hat{X}_{4})}{2},\theta\to \infty \right) = 1-\cos\left( \f{q\pi}{L}(\hat{X}_{1}+\hat{X}_{4}) \right).
\end{equation}
Solving the above condition, we get
\begin{equation}
	d_{41,\theta\gg 1} = -i \f{L}{\pi q } \log\left[ \f{ \f{q\beta}{2L}\left(1-f_{\theta\to \infty,X_{1,4}} \right)\log 2 + \sqrt{ 1-f_{\theta\to \infty,X_{1,4}} \left( f_{\theta\to \infty,X_{1,4}}  -2 \right) \left( \f{q\beta}{2L}\log 2 \right)^{2} }  }{ i+ \f{q\beta}{2L}\log 2 } \right].
\end{equation}
By noting that we are focusing on the high temperature regime, $L/\beta \gg 1$ with keeping $q$ constant not promotional to $L/\beta $, i.e.,
\begin{equation}
	\f{L}{q\beta} \gg 1
\end{equation}
the above result reduces to,
\begin{equation}\label{eq:correlaLengthSuperHigh}
	\begin{aligned}
		d_{41,\theta\gg 1} &\approx \f{\beta}{2\pi} f_{\theta\to \infty,X_{1,4}} \log 2\\
		& = \f{\beta}{2\pi} \cdot  f\left( \f{(\hat{X}_{1}+\hat{X}_{4})}{2},\theta\to \infty \right) \log 2.
	\end{aligned}
\end{equation}
In terms of this correlation length, the mutual information can be written as
\begin{equation}
	\begin{aligned}
		&\left.I(A_{1},A_{1}')\right|_{\text{High temperature limit w/ } \theta\gg 1}\\
		&\approx \begin{dcases}
			0  &   d_{41,\theta\gg 1} \leq \hat{X}_{4}-\hat{X}_{1},\\
			-\frac{c}{3} \log\left[\exp\left[\f{2L}{q\beta} \cdot \f{\sin\left( \f{q\pi}{L}(\hat{X}_{4}-\hat{X}_{1}) \right) }{\cos\left( \f{q\pi}{L}(\hat{X}_{4}-\hat{X}_{1}) \right) - \cos\left( \f{q\pi}{L}(\hat{X}_{1}+\hat{X}_{4}) \right)} \right] -1 \right] 	&  \hat{X}_{4}-\hat{X}_{1} {<}  d_{41,\theta\gg 1}.
		\end{dcases}
	\end{aligned}
\end{equation}
This result is valid for the high temperature region,  $L/(q\beta) \gg 1$, and the large $\theta$ regime which validates the approximation \eqref{eq:AngleExpanLargeTheta} with the condition \eqref{eq:approConditionSuperHigh} .

Using the above result, let us check whether there is a correlation between two intervals, which are distant but located between $x= \f{(n+1)L}{q}$ and $\f{nL}{q}$.
For simplicity, let us assume that one of two intervals is located around the middle of the fixed points, and the other is located near either $\f{(n+1)L}{q}$ or $\f{nL}{q}$.
Since, under this assumption,
the distance between the two points, $\hat{X}_{1},\hat{X}_{4}$ is approximately given by $\hat{X}_{4}-\hat{X}_{1}\sim \f{L}{2q}$, we obtain the inequality about $\hat{X}_4-\hat{X}_1$,
\begin{equation}
	\hat{X}_{4}-\hat{X}_{1}\sim \f{L}{2q} \gtrsim d_{41,\theta\gg 1}.
\end{equation}
Thus, there is no non-local correlation between the two intervals considered above.
More generally, there would be no non-local correlation between two distant intervals whose distance is the order of $L/q$.  

\subsection{Entanglement dynamics under the small interval limit \texorpdfstring{$\f{L}{q} \gg |\hat{X}_{i}-\hat{X}_{j}| $}{L/q} \label{subsec:smallIntEnt}}

So far, we have considered the large $\theta$ limit to simplify the expressions of the entanglement entropy.
In this section, we will consider another limit to simplify the expression, the so-called small interval limit,$\f{L}{q} \gg |\hat{X}_{i}-\hat{X}_{j}| $.

In this small interval limit, moreover, let us impose the following condition on the size of the interval, $|\hat{X}_{i}-\hat{X}_{j}|$ \footnote{The condition can be also written as,
\begin{equation}
	\begin{aligned}
			1 &\gg  \dfrac{  \left|\hat{X}_{i}-\hat{X}_{j}\right| }{   \dfrac{ L_{\text{eff}} }{q}\cdot \left[  1 -\tanh(2\theta) \cos\left( \frac{2\pi q}{L}\cdot \frac{ \hat{X}_{i}+\hat{X}_{j} }{2} \right)  \right]  } \pi = \f{ \left|\hat{X}_{i}-\hat{X}_{j}\right| }{ \dfrac{ L_{\text{eff}} }{q} f\left(\frac{ \hat{X}_{i}+\hat{X}_{j} }{2},\theta\right) },
	\end{aligned}
\end{equation}
where $f(x,\theta)$ is the envelop function \eqref{eq:enveFunc}.
 },
\begin{equation}\label{eq:smallIntCondi}
	\begin{aligned}
			1 &\gg  \f{  \f{\pi q}{L}\left|\hat{X}_{i}-\hat{X}_{j}\right| }{ \cosh (2\theta) -\sinh(2\theta) \cos \left(  \f{\pi q}{L}\left(\hat{X}_{i}+\hat{X}_{j}\right) \right) }.
	\end{aligned}
\end{equation}
Then, by the next-to-leading order in this expansion, $\varphi_i-\varphi_j$ is approximately given by
\begin{equation}\label{eq:smallInterApprox}
	\begin{aligned}
		\varphi_i-\varphi_j &\approx (m_{i}-m_{j})\pi +  \f{  \f{\pi q}{L}\left(\hat{X}_{i}-\hat{X}_{j}\right) }{ \cosh (2\theta) -\sinh(2\theta) \cos \left(  \f{\pi q}{L}\left(\hat{X}_{i}+\hat{X}_{j}\right) \right) }. 
	\end{aligned} 
\end{equation}
In addition, in this limit, the universal part of the entanglement entropy, i.e., the first term of \eqref{eq:EEU+NU}, also can reduce to\footnote{We note that, under the small interval limit, the universal part can be also written as
\begin{equation}
	\begin{aligned}
		S_{A_{i}}^{\text{U.}} &\approx \frac{c}{3} \log \left[ 1 -\tanh(2\theta) \cos\left( \frac{\pi q}{L}\left(\hat{X}_{1}+\hat{X}_{2}\right)\right) \right]\\ 
		& = \f{c}{3}\log \left[ f\left( \frac{\hat{X}_{1}+\hat{X}_{2}}{2},\theta \right)  \right],
	\end{aligned}
\end{equation}
where $f(x,\theta)$ is the envelop function \eqref{eq:enveFunc}.
}
\begin{equation}\label{eq:uniApprox}
	\begin{aligned}
		S_{A_{i}}^{\text{U.}} &= \f{c}{6}\log{\left[\prod_{i=1,2}\left(1-\tanh{2\theta}\cos{\left(\f{2q \pi X_i}{L}\right)}\right)\right]}\\
		 &\approx \frac{c}{3} \log \left[ 1 -\tanh(2\theta) \cos\left( \frac{\pi q}{L}\left(\hat{X}_{1}+\hat{X}_{2}\right)\right) \right].
	\end{aligned}
\end{equation}

In the following, we consider the behavior of the entanglement entropy in the low and high effective regions.

\paragraph{Low effective temperature region}
In the low effective temperature region, $\tau_{\text{Mod.}}<1$, the theory-dependent pieces of the entanglement entropy for $A_1$ and $A_2$, \eqref{eq:EEvac}, are approximately given by
\begin{equation}
	\begin{aligned}
		 S_{A_{1}}^{\text{T.D.}} &\approx \f{c}{3}\log{\left(\f{L_{\text{eff}}}{\pi \epsilon}\right)}+\f{c}{3}\log{ \left(  \f{ \dfrac{\pi }{L}\left(\hat{X}_{1}-\hat{X}_{2}\right) }{ \cosh (2\theta) -\sinh(2\theta) \cos \left(  \dfrac{\pi q}{L}\left(\hat{X}_{1}+\hat{X}_{2}\right) \right) }  \right) }\\
		 & = \f{c}{3}\log{ \left(  \f{  \dfrac{1}{\epsilon} \left(\hat{X}_{1}-\hat{X}_{2}\right) }{ 1 -\tanh (2\theta) \cos \left(  \dfrac{\pi q}{L}\left(\hat{X}_{1}+\hat{X}_{2}\right) \right) }  \right) },
	\end{aligned}\label{eq:EELowNUA1smallInter}
\end{equation}
\begin{equation}
	\begin{aligned}
		S_{A_{2}}^{\text{T.D.}} &\approx \f{c}{3}\log{\left(\f{L_{\text{eff}}}{\pi \epsilon}\right)}+\f{c}{3}\log{   \sin \left( \f{l}{q} \pi \right)  }.
	\end{aligned}\label{eq:EELowNUA2smallInter}
\end{equation}
Then, combining the universal part \eqref{eq:uniApprox} with \eqref{eq:EELowNUA1smallInter} and \eqref{eq:EELowNUA2smallInter}, the entanglement entropies in this region are approximately given by
\begin{equation}
	S_{A_{1}} \approx \f{c}{3} \log\left( \dfrac{ \hat{X}_{1}-\hat{X}_{2} }{\epsilon} \right),
\end{equation}
\begin{equation}
	S_{A_{2}} \approx \f{c}{3} \log \left( \f{L}{\pi \epsilon} \left[\cosh(2\theta) -\sinh(2\theta) \cos\left( \frac{\pi q}{L}\left(\hat{X}_{1}+\hat{X}_{2}\right) \right) \right]  \right) + \f{c}{3}\log{   \sin \left( \f{l}{q} \pi \right)  }.
\end{equation}

\paragraph{High effective temperature region}
Now, we focus on the behavior of $S_{A_1}$ and $S_{A_2}$ in the high effective temperature region, $\tau_{\text{Mod.}}>1$. In this region of the limit considered here, using the approximation \eqref{eq:smallInterApprox}, we can evaluate the theory-dependent parts \eqref{eq:theEE} as
\begin{equation}
	\begin{aligned}
		 S_{A_{1}}^{\text{T.D.}} \approx \f{c}{3}\log{\left(\f{ \beta }{\pi \epsilon}\right)} +  \f{c}{3}\log\left[ \sinh \left( \f{\pi }{\beta}\cdot \f{ \hat{X}_{1}-\hat{X}_{2} }{ 1 -\tanh(2\theta) \cos\left( \frac{\pi q}{L}\left(\hat{X}_{1}+\hat{X}_{2}\right) \right) } \right) \right],
	\end{aligned}\label{eq:EEHighNUA1smallInter}
\end{equation}
\begin{equation}
	\begin{aligned}
		S_{A_{2}}^{\text{T.D.}} &\approx \f{c}{3}\log{\left(\f{ \beta }{\pi \epsilon}\right)}+\f{c}{3}\log{  \left[  \sinh \left( \f{L \cosh (2\theta) }{\beta}\cdot \f{l}{q}\pi  \right)   \right]   }.
	\end{aligned}\label{eq:EEHighNUA2smallInter}
\end{equation}

Consequently, the entanglement entropies for $A_{i=1,2}$ reduce to
\begin{equation}
	\begin{aligned}
		 S_{A_{1}} & \approx  \f{c}{3}\log{\left(\f{ \beta  }{\pi \epsilon} \cdot \left[ 1 -\tanh(2\theta) \cos\left( \frac{\pi q}{L}\left(\hat{X}_{1}+\hat{X}_{2}\right)\right) \right] \right)}\\
		 & \qquad  +  \f{c}{3}\log\left[ \sinh \left( \f{\pi }{\beta}\cdot \f{ \hat{X}_{1}-\hat{X}_{2} }{ 1 -\tanh(2\theta) \cos\left( \frac{\pi q}{L}\left(\hat{X}_{1}+\hat{X}_{2}\right) \right) } \right) \right],
	\end{aligned}
\end{equation}
\begin{equation}
	\begin{aligned}
		S_{A_{2}} &\approx \f{c}{3}\log{\left(\f{ \beta  }{\pi \epsilon} \cdot \left[ 1 -\tanh(2\theta) \cos\left( \frac{\pi q}{L}\left(\hat{X}_{1}+\hat{X}_{2}\right)\right) \right] \right)}\\		& \qquad +\f{c}{3}\log{  \left[  \sinh \left( \f{L \cosh (2\theta) }{\beta}\cdot \f{l}{q}\pi  \right)   \right]   }.
	\end{aligned}
\end{equation}

\subsubsection*{Mutual information in the small interval limit}
We will explore the $\theta$ dependence of the mutual information in the small interval limit without the large $\theta$ expansion. 
The mutual information considered here is between $A_1$ and $A'_1$.
Here, the distance between two interval is also assumed to be small compared to $L/q$. 


We start with the low effective temperature regime, $\tau_{\text{Mod.}}=L_{\text{eff}}/\beta \leq 1$. For this parameter regime, the mutual information $I(A_{1},A_{1}')$ is given by \eqref{eq:MILowGeneral}. 
We consider the behavior of the mutual information in \eqref{eq:smallIntCondi}, and then use the approximation of $\varphi_i-\varphi_j$ in \eqref{eq:smallInterApprox}. The function, $\sin\left( \f{\varphi_{i}-\varphi_{j}}{q} \right)$, reduces to
\begin{equation}
	\begin{aligned}
		\sin\left( \f{\varphi_{i}-\varphi_{j}}{q} \right) &\approx \f{  \f{\pi }{L}\left(\hat{X}_{i}-\hat{X}_{j}\right) }{ \cosh (2\theta) -\sinh(2\theta) \cos \left(  \f{\pi q}{L}\left(\hat{X}_{i}+\hat{X}_{j}\right)\right) }\\
		&\approx \f{  \f{\pi }{L_{\text{eff}}}\left(\hat{X}_{i}-\hat{X}_{j}\right) }{ f\left( \f{\hat{X}_{i}+\hat{X}_{j}}{2},\theta \right) },
	\end{aligned}
\end{equation}
where $f(x,\theta)$ is the envelop function in \eqref{eq:enveFunc}.
Consequently, the ratio determining the behavior of the mutual information in \eqref{eq:MILowGeneral} becomes
\begin{equation}
	\begin{aligned}
		\frac{ \sin \left( \frac{\varphi_{1}-\varphi_{2}}{q} \right) \sin \left( \frac{\varphi_{3}-\varphi_{4}}{q} \right) }{ \sin \left( \frac{\varphi_{3}-\varphi_{2}}{q} \right) \sin \left( \frac{\varphi_{4}-\varphi_{1}}{q} \right) }
		&\approx \frac{ \sin\left( \frac{\pi}{L}(X_{1}-X_{2}) \right) \sin\left( \frac{\pi}{L}(X_{3}-X_{4}) \right) }{ \sin\left( \frac{\pi}{L}(X_{3}-X_{2}) \right) \sin\left( \frac{\pi}{L}(X_{4}-X_{1}) \right) } \cdot \f{ f\left( \f{\hat{X}_{2}+\hat{X}_{3}}{2},\theta \right) f\left( \f{\hat{X}_{1}+\hat{X}_{4}}{2},\theta \right) }{ f\left( \f{\hat{X}_{1}+\hat{X}_{2}}{2},\theta \right) f\left( \f{\hat{X}_{3}+\hat{X}_{4}}{2},\theta \right) }\\
		&\approx \frac{ \sin\left( \frac{\pi}{L}(X_{1}-X_{2}) \right) \sin\left( \frac{\pi}{L}(X_{3}-X_{4}) \right) }{ \sin\left( \frac{\pi}{L}(X_{3}-X_{2}) \right) \sin\left( \frac{\pi}{L}(X_{4}-X_{1}) \right) },
	\end{aligned}
\end{equation}
where, in the final line, we used the following approximation in the small interval limit
\begin{equation}
	f\left( \f{\hat{X}_{2}+\hat{X}_{3}}{2},\theta \right) f\left( \f{\hat{X}_{1}+\hat{X}_{4}}{2},\theta \right) \approx f\left( \f{\hat{X}_{1}+\hat{X}_{2}}{2},\theta \right) f\left( \f{\hat{X}_{3}+\hat{X}_{4}}{2},\theta \right).
\end{equation}
Thus, since in the small interval limit, the envelop functions are cancelled out, the mutual information $I(A_{1},A_{1}')$  is reduced to the conventional vacuum one,
\begin{equation}\label{eq:MIsmallLow}
	I_{\text{Low}}(A_{1},A_{1}') \approx \max\left\{ 0, \frac{c}{3} \log\left[ \frac{ \sin\left( \frac{\pi}{L}(X_{1}-X_{2}) \right) \sin\left( \frac{\pi}{L}(X_{3}-X_{4}) \right) }{ \sin\left( \frac{\pi}{L}(X_{3}-X_{2}) \right) \sin\left( \frac{\pi}{L}(X_{4}-X_{1}) \right) }\right] \right\}.
\end{equation}
{Next, we consider the} effective temperature regime.
In particular, for simplicity, we take the high effective temperature and high temperature limits, $\tau_{\text{Mod.}}=L_{\text{eff}}/\beta,L/\beta \gg 1$  and then evaluate the mutual information in \eqref{eq:mutualHigh}. 
In this limit, the ratio in \eqref{eq:ratioHigh} approximately reduces to
\begin{equation}
	\begin{aligned}
		\frac{ \sinh{\left[\f{L_{\text{eff}}\left(\varphi_1-\varphi_2\right)}{q\beta}\right]}  \sinh{\left[\f{L_{\text{eff}}\left(\varphi_3-\varphi_4\right)}{q\beta}\right]} }{ \sinh{\left[\f{L_{\text{eff}}\left(\varphi_3-\varphi_2\right)}{q\beta}\right]} \sinh{\left[\f{L_{\text{eff}}\left(\varphi_4-\varphi_1\right)}{q\beta}\right]} } \approx \f{1}{\exp\left[\f{2L_{\text{eff}}\left(\varphi_4-\varphi_1\right)}{q\beta}\right]-1},
	\end{aligned}
\end{equation}
where we assume that
\begin{equation}\label{eq:MIratioAppCond}
	\f{L_{\text{eff}}\left(\varphi_1-\varphi_2\right)}{q\beta},\f{L_{\text{eff}}\left(\varphi_3-\varphi_4\right)}{q\beta},\f{L_{\text{eff}}\left(\varphi_3-\varphi_2\right)}{q\beta} \gg 1.
\end{equation}


Then, by assuming that the two intervals satisfy the adjacent condition \eqref{eq:smallIntCondi}, i.e.,
\begin{equation*}
   \f{  \f{\pi q}{L}\left|\hat{X}_{4}-\hat{X}_{1}\right| }{ \cosh (2\theta) -\sinh(2\theta) \cos \left(  \f{\pi q}{L}\left(\hat{X}_{1}+\hat{X}_{4}\right) \right) } \ll 1,
\end{equation*}
and using the approximation \eqref{eq:smallInterApprox}, the ratio approximately reduces to
\begin{equation}
	\begin{aligned}
		\frac{ \sinh{\left[\f{L_{\text{eff}}\left(\varphi_1-\varphi_2\right)}{q\beta}\right]}  \sinh{\left[\f{L_{\text{eff}}\left(\varphi_3-\varphi_4\right)}{q\beta}\right]} }{ \sinh{\left[\f{L_{\text{eff}}\left(\varphi_3-\varphi_2\right)}{q\beta}\right]} \sinh{\left[\f{L_{\text{eff}}\left(\varphi_4-\varphi_1\right)}{q\beta}\right]} } &\approx \f{1}{\exp\left[  \f{2\pi }{\beta}\cdot \f{ \hat{X}_{4}-\hat{X}_{1} }{ 1 -\tanh(2\theta) \cos\left( \frac{\pi q}{L}\left(\hat{X}_{1}+\hat{X}_{4}\right) \right) }  \right]-1}\\
		& = \f{1}{\exp\left[  \frac{2\pi}{\beta}\cdot \frac{ \hat{X}_{4}-\hat{X}_{1} }{ f\left( \f{\hat{X}_{1}+\hat{X}_{4}}{2},\theta \right) }   \right]-1}.
	\end{aligned}
\end{equation}
Consequently, in this limit, the behavior of the mutual information is determined by  
\begin{equation}\label{eq:MIsmallInterHigh}
	I_{\text{High}}(A_{1},A_{1}') \approx  \max\left\{ 0, -\frac{c}{3} \log\left[ \exp\left(  \frac{2\pi}{\beta}\cdot \frac{ \hat{X}_{4}-\hat{X}_{1} }{ f\left( \f{\hat{X}_{1}+\hat{X}_{4}}{2},\theta \right) }   \right)-1 \right] \right\}.
\end{equation}
We note that this result is valid under the condition \eqref{eq:smallIntCondi}.

From the above expression \eqref{eq:MIsmallInterHigh}, we can read off the correlation length, $d_{41}$, which corresponds to the thermal correlation length. The correlation length can be determined by
\begin{equation}
	\frac{c}{3} \log\left[ \exp\left(  \frac{2\pi}{\beta}\cdot \frac{ d_{41} }{ f\left( \f{\hat{X}_{1}+\hat{X}_{4}}{2},\theta \right) }   \right)-1 \right] =0,
\end{equation}
giving 
\begin{equation}\label{eq:themalCorreLenSma}
	\begin{aligned}
		d_{41}&=\f{\beta}{2\pi} \cdot  f\left( \f{\hat{X}_{1}+\hat{X}_{4}}{2},\theta \right)\log 2 =\f{\beta_{\text{eff}}\left( \f{\hat{X}_{1}+\hat{X}_{4}}{2},\theta  \right) }{2\pi}\log 2,
	\end{aligned}
\end{equation}
where we defined the effective inverse temperature $\beta_{\text{eff}}(x,\theta)$ by
\begin{equation}\label{eq:effLocalTemp}
	\beta_{\text{eff}}(x,\theta) \coloneqq \beta \,  f\left( x,\theta \right)=\beta \left[ 1-\tanh( 2\theta) \cos \left( \f{2\pi q}{L}x \right) \right]. 
\end{equation}
This correlation length is the same as that in the limit, where $L_{\text{eff}}/\beta,L/\beta,\theta \gg 1$, \eqref{eq:correlaLengthSuperHigh}.

In terms of the above correlation length, the mutual information can be written as
\begin{equation} \label{eq:MI-in-high-high-large-limit}
	I_{\text{High}}(A_{1},A_{1}') \approx  \max\left\{ 0, -\frac{c}{3} \log\left[ \exp\left( \frac{ \hat{X}_{4}-\hat{X}_{1} }{ d_{41} } \log 2 \right)-1 \right] \right\}.
\end{equation}
We can see from (\ref{eq:MI-in-high-high-large-limit}) that when two intervals are separated by more than the thermal correlation length, the mutual information vanishes, while when they are separated by less than the length, the mutual information takes the non-trivial value,
\begin{equation}
	\begin{aligned}
		I_{\text{High}}(A_{1},A_{1}') &\approx
		\begin{dcases}
	0 & \quad d_{41} < \hat{X}_{4}-\hat{X}_{1},\\
	-\frac{c}{3} \log\left[ \exp\left( \frac{ \hat{X}_{4}-\hat{X}_{1} }{ d_{41} } \log 2 \right)-1 \right]  & \quad  0<  \hat{X}_{4}-\hat{X}_{1} \leq d_{41}.
		\end{dcases}
	\end{aligned}
\end{equation} 

\begin{figure}[H]
\begin{center}
\includegraphics[scale=0.9]{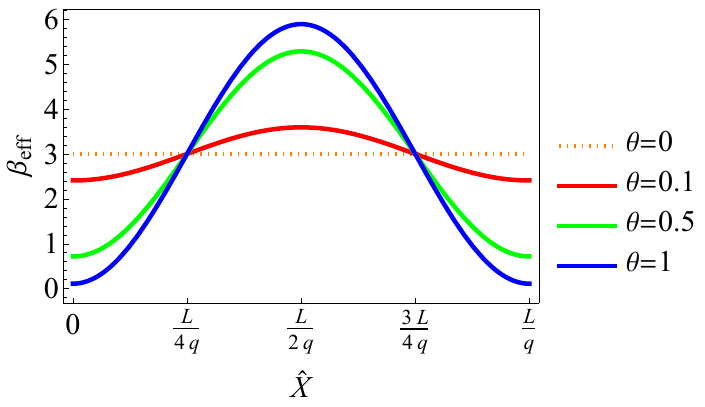}
\caption{Temperature distribution, $\beta_{\text{eff}}(\hat{X},\theta)$, (\ref{eq:effLocalTemp}). The original inverse temperature is chosen to be $\beta=3$, which corresponds to the $\theta=0$ case. }
\label{fig:TemperatureDistribution}
\end{center}
\end{figure}

The crucial key ingredient of the above result is the position-dependence of the thermal correlation length, $d_{41}$, or equivalently the effective inverse temperature $\beta_{\text{eff}}(x,\theta)$. The envelop function controls the position dependence of the effective temperature:
when $\f{\hat{X}_{1}+\hat{X}_{4}}{2}=\f{L}{2q}$,  the effective inverse temperature takes the maximal value, $\beta_{\text{eff}.\max}=\beta(1+\tanh(2\theta))$; when $\f{\hat{X}_{1}+\hat{X}_{4}}{2}=\f{L}{4q},\f{3L}{4q}$, the effective temperature reduces to the original temperature, $\beta$; and as the point $\f{\hat{X}_{1}+\hat{X}_{4}}{2}$ approaches $0$ or $\f{L}{q}$ while keeping the condition \eqref{eq:MIratioAppCond}
, the temperature goes to the minimum value $\beta_{\text{eff}.\min}=\beta(1-\tanh(2\theta))$,
 \begin{equation}
 	\begin{aligned}
 		\beta_{\text{eff}}\left( \f{\hat{X}_{1}+\hat{X}_{4}}{2},\theta  \right)& \to 
 		\begin{dcases}
 			\beta_{\text{eff}.\max}=\beta(1+\tanh(2\theta)) &  \text{ as }\f{\hat{X}_{1}+\hat{X}_{4}}{2} \to \f{L}{2q},\\
 			\beta & \text{ as }\f{\hat{X}_{1}+\hat{X}_{4}}{2} \to \f{L}{4q}  \text{ or } \f{3L}{4q}, \\
 			\beta_{\text{eff}.\min}=\beta(1-\tanh(2\theta)) & \text{ as }\f{\hat{X}_{1}+\hat{X}_{4}}{2} \to 0 \text{ or } \f{L}{q}.
 		\end{dcases}
 	\end{aligned}
 \end{equation}
 See Fig. \ref{fig:TemperatureDistribution} for the local inverse temperature distribution, $\beta_{\text{eff}}$. Thus, the closer the location of the subsystem is to $x=\f{mL}{q}$, where $m$ runs from $0$ to $q-1$, the higher the local temperature is. The local temperature is maximized at $x=\f{mL}{q}$, while it is minimized at $x=\f{L}{2q}+\f{mL}{q}$.
 
One would be able to relate this behavior to the geometric property of the CFTs curved geometry, e.g., the curvature in \eqref{eq:RicciCurvature}, and also the bulk geometry discussed in Section \ref{sec:gravitydual}.


\if[0]
\sout{We note that the above result is valid for the case where}  \textcolor{red}{\sout{an unstable fixed point is not included between the two small intervals. For the case where unstable fixed points are not included between them, we need some modifications.}} \textcolor{blue}{\bf $\rightarrow$ MN: These two sentences look to be inconsistent with each other. Also, unstable point should be replaced with $\f{mL}{q}$ because this point has not been defined yet.} \textcolor{magenta}{\bf AM: It seems that the treatment of the situation would require more discussions, so I remove this paragraph here.}
 \sout{Let us slightly discuss the case. In this case, since $m_{4}-m_{1}=l \in \mathbb{N}$ \textcolor{blue}{$\leftarrow \mathbb{Z}$?, also $m_i$ has not been defined yet, I think.} in the approximation \eqref{eq:smallIntCondi}, $\varphi_4-\varphi_1$ should be approximated by $l\pi $. Here, the condition, where the approximation is valid, is}
\begin{equation}
	\begin{aligned}
			1 &\gg  \f{  \f{\pi q}{L}\left|\hat{X}_{1}-\hat{X}_{4}\right| }{ \cosh (2\theta) -\sinh(2\theta) \cos \left(  \f{\pi q}{L}\left(\hat{X}_{1}+\hat{X}_{4}\right) \right) } \approx \f{\pi}{e^{2\theta}}
	\end{aligned}
\end{equation}
\sout{where we used $\hat{X}_{1}-\hat{X}_{4}\approx \f{L}{q}$ and $\f{\hat{X}_{1}+\hat{X}_{4}}{2} \approx \f{L}{2q}$.
 Then, the mutual information reduces to}
\begin{equation}
	\begin{aligned}
		I(A_{1},A_{1}')|_{m_{4}-m_{1}=l} &\approx  \max\left\{ 0, -\frac{c}{3} \log\left[ \exp\left(\f{2\pi l  L_{\text{eff}}}{q\beta}\right)-1 \right] \right\}=0.
	\end{aligned}
\end{equation}
\sout{We can see from this equation that there are no correlations between two regions satisfying $m_{4}-m_{1}=l\neq 0$.
The behaviors, described above, of $I(A_1,A'_1)$ are significantly different from a correlation length defined through the mutual information for the low temperature case \eqref{eq:MIsmallLow}: the correlation length is just given by the vacuum one, and it does not have a position-dependence. }

\fi

\subsection{Entropy gap under the small interval limit}


Using the entanglement entropies in the small interval limit obtained in Section \ref{subsec:smallIntEnt}, let us re-investigate the behaviors of the entropy gap, which we obtained in Section \ref{sec:entgap}. 
In obtaining the entropy gap in Section \ref{sec:entgap}, we considered the parameter region, where $L/\beta \ll 1, \theta \gg 1$ with $L \cosh (2\theta) /\beta=1$.
The small interval limit does not need to be in this parameter region. 

First, we focus on the region $A_{1}$. In this case, by using the expressions in \eqref{eq:EELowNUA1smallInter} and \eqref{eq:EEHighNUA1smallInter} and the condition $L \cosh (2\theta) /\beta=1$, we can see that the entropy gap vanishes,
\begin{equation*}
	\Delta S_{A_{1}}=0.
\end{equation*}
This result, which is obtained without taking the large $\theta$ limit, is consistent with the previous result in \eqref{eq:EEGapA1Limit}.
Then, we consider the other region, $A_{2}$. For this region, we can use the expressions in \eqref{eq:EELowNUA2smallInter} and \eqref{eq:EEHighNUA2smallInter}, and see that the entropy gap for this region is equal to the previous one,
\begin{equation*}
	\Delta S_{A_{2}}= \f{c}{3}  \text{Min} \left\{ \log{ \f{ \sinh\left( \dfrac{l}{q}\pi\right) }{\sin\left( \dfrac{l}{q}\pi\right)}  } , \log{ \f{ \sinh\left( \left(1- \dfrac{l}{q} \right)\pi\right) }{\sin\left( \dfrac{l}{q}\pi\right)}  } + \pi  \right\},
\end{equation*}
where we re-added the other entanglement entropy contributions satisfying the homology conditions, and we used $L \cosh (2\theta) /\beta=1$, but did not take the large $\theta$ limit.
These results support the existence of the entropy gap.
One would be able to relate this behavior to the geometric property of the CFTs curved geometry, e.g., the curvature in \eqref{eq:RicciCurvature}, and also the bulk geometry discussed in Section \ref{sec:gravitydual}.

\subsection{Emergence of the Vacuum entanglement structure}
Now, we will explain why the entanglement structure, resembling the vacuum one in the periodic system with $L/q$, emerges in the low temperature and large $\theta$ limit, $\beta/L\gg 1$, and $\theta \gg 1$.
We begin by considering the thermal state in the low temperature limit.
At the leading order in the $\beta/L \gg 1$ limit, the thermal state considered approximately reduces to the density operator for the vacuum state,
\be
\f{e^{-\beta H_{\text{q-M\"obius}}}}{\Tr e^{-\beta H_{\text{q-M\"obius}}}} \underset{\beta/L \gg 1} {\approx}\ket{0}\bra{0},
\ee
where $\ket{0}$ denotes the vacuum state of $H_{\text{q-M\"obius}}$.
Subsequently, we closely look at the entanglement structure in the large $\theta$ limit.
At the leading order in the large $\theta$ expansion, except for $x=\f{mL}{q}$, $H_{\text{q-M\"obius}}$ reduces to 
\be \label{eq:approximation-of-H}
\begin{split}
    H_{\text{q-M\"obius}} &\approx 2\int^L_0 dx \sin^2{\left(\f{q\pi x}{L}\right)}\left(T(x)+\overline{T}(x)\right)=2\sum_{m=0}^{q-1}\int^{\f{(m+1)L}{q}}_{\f{mL}{q}}dx \sin^2{\left(\f{q\pi x}{L}\right)}\left(T(x)+\overline{T}(x)\right)\\
    &=\sum^{q-1}_{m=0}{H_{\text{SSD},L/q,m}},
\end{split}
\ee
where $H_{\text{SSD},L/q,m}$ is the SSD Hamiltonian acting on the spatial interval between $x=m L/q$ and $(m+1)L/q$. The length of this interval is $L/q$.
Therefore, in the large limit the vacuum state for $H_{\text{q-M\"obius}}$ may approximately reduce to 
\be \label{eq:approximation-of-V}
\ket{0} \approx \prod_{m=0}^{q-1}\ket{0,m}+ \cdots
\ee
where $\ket{0,m}$ is the vacuum state of ${H_{\text{SSD},L/q,m}}$, and {$\cdots$ means the higher-order contributions.}
As in \cite{Wen:2018vux}, the vacuum state of {$H_{\text{SSD},L/q,m}$} on the strip is the same as that for the uniform Hamiltonian with the periodic boundary condition.
Thus, in the low temperature and large $\theta$ limit, the entanglement entropy for the subsystems, not including any $\f{mL}{q}$, is approximated by the vacuum one for the spatially-periodic system with the size of $L/q$. 
The entanglement entropy for the subsystem, including some of $\f{mL}{q}$, does not follow the approximations (\ref{eq:approximation-of-H}) and (\ref{eq:approximation-of-V}).
In this case, we need to carefully treat the higher-order terms in the large $\theta$ expansion of $H_{\text{q-M\"obius}}$ in (\ref{eq:approximation-of-H}).
\section{Gravity dual}\label{sec:gravitydual}

So far, we have presented the behavior of entanglement entropy and mutual information in $2$d holographic CFTs on the curved geometry in (\ref{eq:background}).
In this section, we will consider its bulk dual. 
To this end, we start with the standard BTZ metric~\cite{Banados:1992wn}
\begin{equation}
	ds^{2}_{\text{bulk}} = \f{dr^{2}}{r^{2}-r_{0}^{2}} - r^{2} \left( \f{ d\xi_{\tau} -d\bar{\xi}_{\tau} }{2} \right)^{2} + (r^{2}-r_{0}^{2} ) \left( \f{ d\xi_{\tau} +d\bar{\xi}_{\tau} }{2} \right)^{2},
\end{equation}
where $r_{0}$ is the radius of a black hole horizon, and we introduce the new coordinates, 
$(\xi_{\tau},\bar{\xi}_{\tau})=(\xi(w)+\tau,\bar{\xi}(\bar{w})+\tau)$. Here, $\xi,\bar{\xi}$ are given by \eqref{eq:xi}. 
\if[0]
\textcolor{red}{\sout{, and we introduce the Euclidean time $\tau$ as the parameter of the trivial (Euclidean) $q$-M\"obius time evolution,}
\begin{equation}
	e^{-\tau H_{\text{q-M\"obius}}}\, \rho \, e^{\tau H_{\text{q-M\"obius}}} =\rho.
\end{equation}
\sout{(The above metric would correspond to the state,}
\begin{equation}
	e^{-\tau H_{\text{q-M\"obius}}}\,U[\xi]\rho[X]U[\xi]^{\dagger}\, e^{\tau H_{\text{q-M\"obius}}} =e^{-\tau H_{\text{q-M\"obius}}} \rho[\xi(X)]  e^{\tau H_{\text{q-M\"obius}}} =\rho[\xi(X),\tau],
\end{equation}
\sout{where $U[\xi]$ is a unitary transformation inducing the conformal transformation from $w$ to $\xi$.)}
} \textcolor{blue}{\bf $\leftarrow$ MN: What you want to say here? Could you let me know about it? } \textcolor{magenta}{AM: I included this statement to explain the difference of the situation here from Appendix B. However, since I have removed Appendix B and we now know the proper construction of the metric we are considering here, I begin to think that this statement is not necessary. So I delete this sentence.}
\fi
Although $\xi$ ($\bar{\xi}$) is a function of $w=T+iX$ ($\bar{w}=T-iX$),  we are interested in the $T=0$ case, since the parameter $T$ is the artificial parameter, not physical one. 
Then, the above BTZ metric becomes the following one,
\begin{equation}\label{eq:BTZmetric}
	ds^{2}_{\text{bulk}} = \f{dr^{2}}{r^{2}-r_{0}^{2}} - r^{2} \left( \f{ d\xi -d\bar{\xi} }{2} \right)^{2} + (r^{2}-r_{0}^{2} ) d\tau^{2}.
\end{equation}
The derivatives at $T=0$, $d\xi,d\bar{\xi}$, are given by
\begin{equation}
	\begin{aligned}
		d\xi = \left. \frac{1}{f\left( -iw ,\theta\right)} dw \right|_{T=0}=\frac{i}{f\left( X ,\theta\right)} dX, \quad  d\bar{\xi} = \left.\frac{1}{f\left( i\bar{w},\theta \right)} d\bar{w}\right|_{T=0}=\frac{-i}{f\left(X,\theta \right)} dX,
	\end{aligned}
\end{equation}
where {$f(x,\theta)$ is given by \eqref{eq:enveFunc}.}
Then, the metric \eqref{eq:BTZmetric} can be written as
\begin{equation}
	ds^{2}_{\text{bulk}} = \f{dr^{2}}{r^{2}-r_{0}^{2}} + (r^{2}-r_{0}^{2} ) d\tau^{2} + \f{ r^{2}}{ f(X,\theta)^{2} } dX^{2}.
\end{equation}
In the large $r$ region, the above metric reduces to the asymptotic metric,
\begin{equation}\label{eq:metric_T=0}
	ds^{2}_{\text{bulk}}|_{r \gg 1} \approx \f{dr^{2}}{r^{2}} + \f{r^{2}}{f(X,\theta)^{2}} \left\{f(X,\theta)^{2} d\tau^{2} +  dX^{2} \right\}.
\end{equation}
Subsequently, we introduce a canonically re-scaled radial coordinate,
\begin{equation}\label{eq:rescalingRadi}
	r'(X,\theta) = \f{r}{f(X,\theta)}.
\end{equation}
We consider the metric in the asymptotic region, $r' \gg 1$, and then it reduces to 
\be
ds^{2}_{\text{bulk}}|_{r' \gg 1} \approx \f{dr'^{2}}{r'^{2}} + r'^{2} \left\{f(X,\theta)^{2} d\tau^{2} +  dX^{2} \right\}
\ee
Thus, the boundary metric is equivalent to (\ref{eq:background}) with the Euclidean signature.

\begin{figure}[H]
\begin{center}
\begin{tabular}{cc}
\subfigure[$q=3$]{
\includegraphics[scale=0.65]{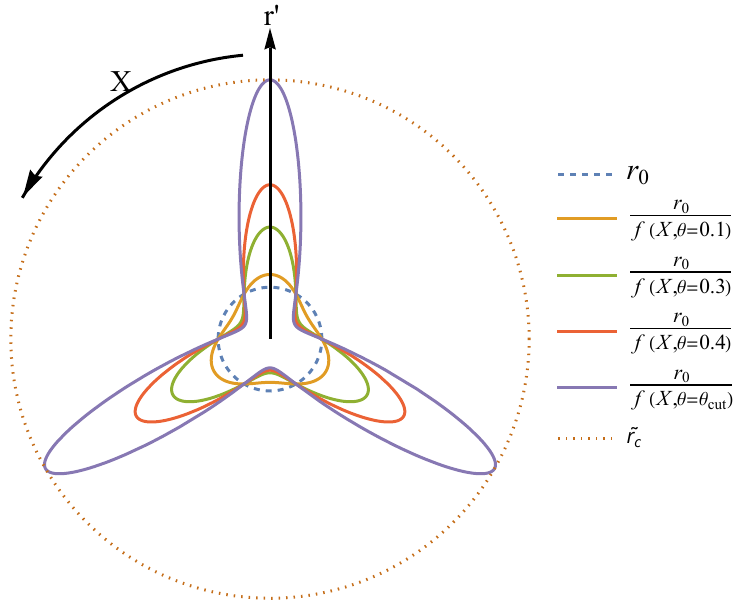}
\label{fig:horizonq=3}
} &
\subfigure[$q=4$]{
\includegraphics[scale=0.65]{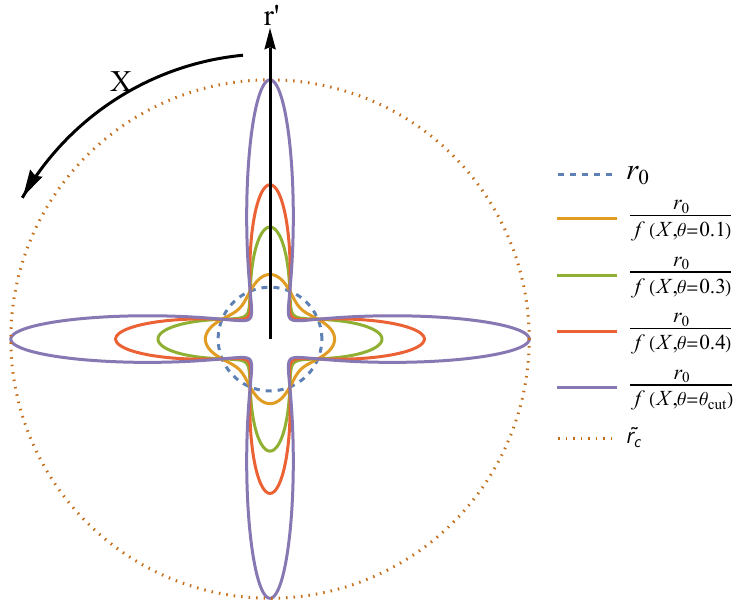}
\label{fig:horizonq=4}
}\\
\subfigure[$q=1$]{
\includegraphics[scale=0.65]{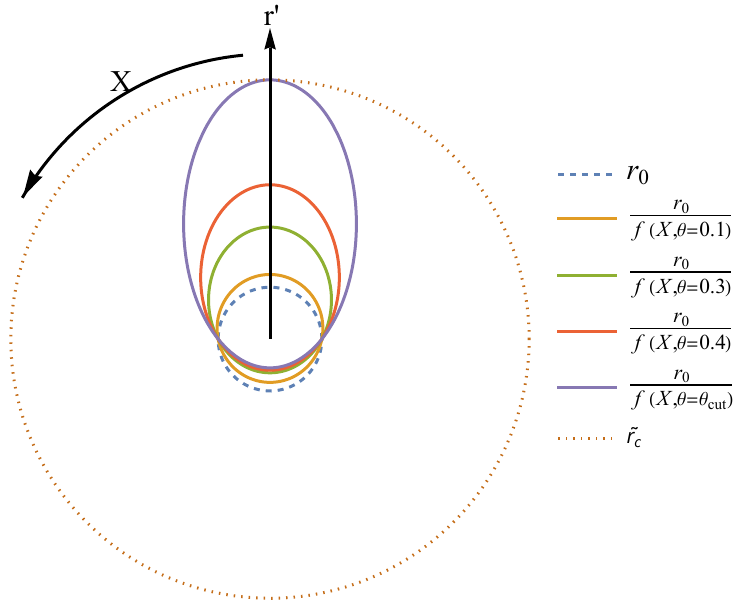}
\label{fig:horizonq=1}
} &
\subfigure[$q=10$]{
\includegraphics[scale=0.65]{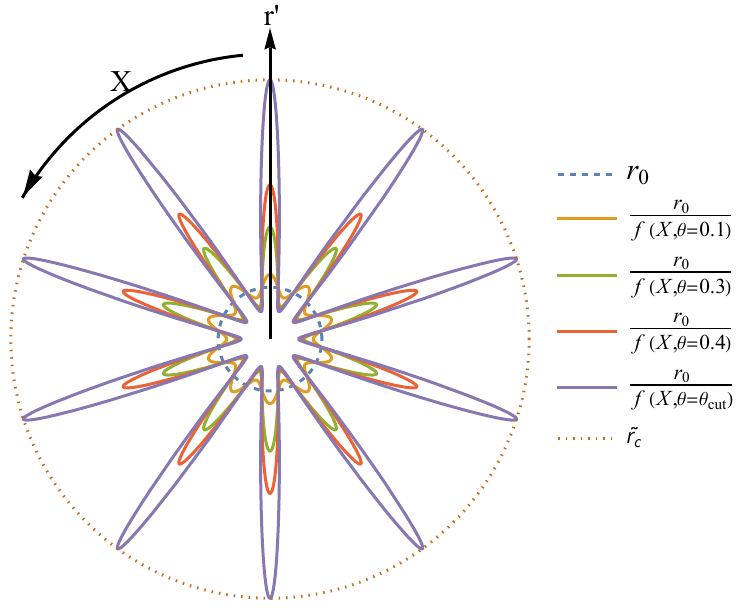}
\label{fig:horizonq=10}
}
\end{tabular}
\caption{Plots of the horizon radius of the black hole (\ref{eq:rescalHorizon}) for various $\theta$ and $q$, and the cutoff surface $\tilde{r}_{\text{c}}$. The dotted lines illustrate the un-deformed horizon, and the dashed lines illustrate the cutoff surface. At $\theta_{\text{cut}}$, the re-scaled horizons touch the cutoff surface. The touching points correspond to $X=\f{mL}{q}$. {Here, we fix the ratio $\tilde{r}_{c}/r_{0}=5$, giving $\theta_{cut}\approx 0.55$.} }
\label{fig:horizons}
\end{center}
\end{figure}

Then, let us consider the radius of the black hole horizon in the new radial coordinate, $r'$.
 In this  radial coordinate, the horizon radius $r'_{0}$ is position-dependent,
\begin{equation}\label{eq:rescalHorizon}
	r'_{0}(X,\theta) = \f{1}{f(X,\theta)}r_{0}.
\end{equation}
In more detail, the horizon radius takes the maximal value at $X=\f{mL}{q}$,
\begin{equation}
	r'_{0,\text{Max}}(X_\text{Un-stable},\theta) = \f{r_{0}}{1-\tanh 2\theta} \geq r_{0}  \quad \text{ at } X_\text{Un-stable}=\f{m L}{q} \quad(m \in \mathbb{Z}),
\end{equation}
while it does the minimal one at $X=\f{L}{2q}+\f{mL}{q}$,
\begin{equation}
	r'_{0,\text{Min}}(X_\text{Stable},\theta) = \f{r_{0}}{1+\tanh 2\theta} \leq r_{0}  \quad \text{ at } X_\text{Stable}=\f{L}{2q}+\f{m L}{q} \quad(m \in \mathbb{Z}).
\end{equation}
In Fig \ref{fig:horizons}, we show the $X$ dependence of $r'_0(X,\theta)$ for the various $\theta$ and $q$.

We note that the re-scaled horizon radius with the sufficiently large $\theta$ becomes comparable to or exceeds a bulk IR cutoff $\tilde{r}_{\text{c}}$, which corresponds to the UV cutoff in the boundary theory.
 For example, if the largest black hole horizon at $X=\f{mL}{q}$ becomes larger than or equal to $\tilde{r}_{\text{c}}$, the bulk description near $X=\f{mL}{q}$ may break down\footnote{{However, our calculations, except for the vicinities of the fixed points, do not show any pathological behaviors even at large $\theta$. This suggests that the boundary cutoff surfaces should be localized on the boundary entangling surfaces rather than spread over the global constant slice $r^\prime=\tilde{r}_c$. See also the similar discussion on the end of the world brane in \cite{2023arXiv231019376N}. }}.
 We can define such a large $\theta_{\text{cut}}$, that leads to the breakdown of the bulk description, as 
 \begin{equation}
 	\tilde{r}_{c} = r'_{0,\text{Max}}(X_\text{Un-stable},\theta_{cut})= \f{r_{0}}{1-\tanh 2\theta_{cut}}.
 \end{equation}
 At the leading order in the large $\theta$ expansion, $\theta_{\text{cut}}$ reduces to
 \begin{equation}
 	\theta_{cut}  \approx \f{1}{4} \log \left[ \f{2\tilde{r}_{c}}{r_{0}} \right].
 \end{equation}

\paragraph{Effective local temperature}

\begin{figure}[H]
\begin{center}
\begin{tabular}{cc}
\subfigure[$\theta=0.1$]{
\includegraphics[scale=0.6]{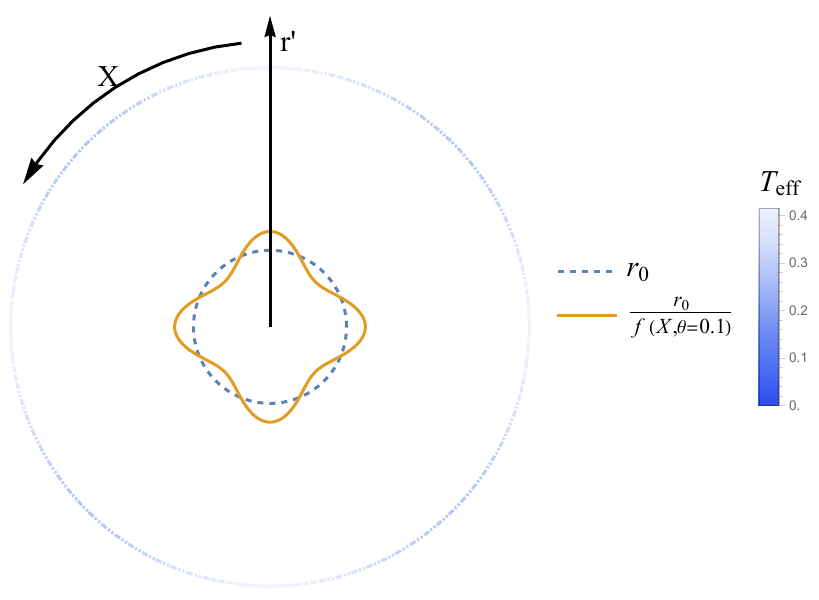}
\label{TempDistvsHorizon1}
} &
\subfigure[$\theta=0.4$]{
\includegraphics[scale=0.6]{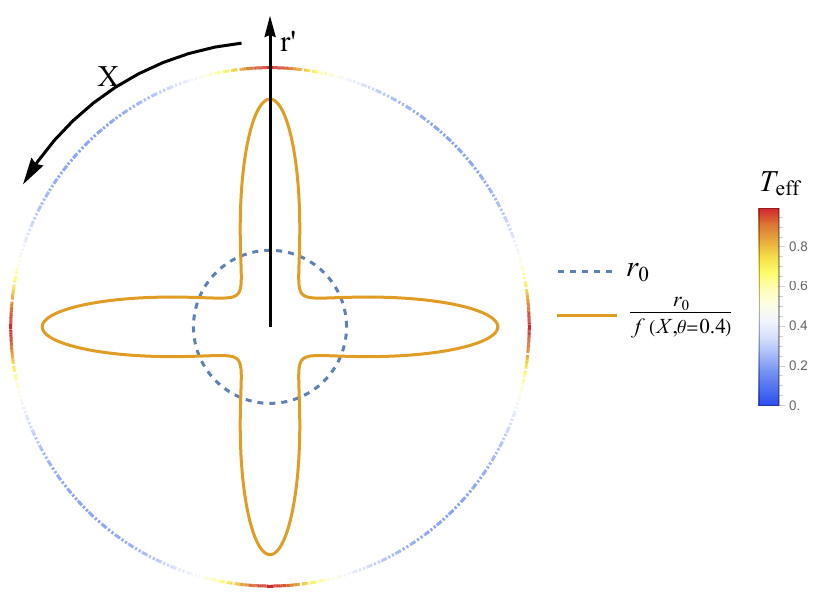}
\label{TempDistvsHorizon2}
}
\end{tabular}
 \centering
\subfigure[$\theta=\theta_{cut}$]{
\includegraphics[scale=0.6]{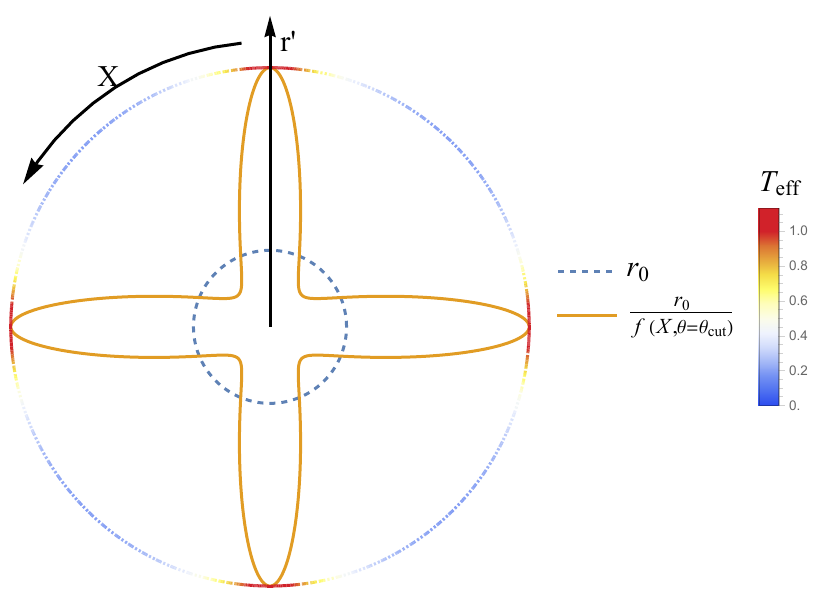}
\label{TempDistvsHorizon3}
}
\caption{Plots of the black hole horizon radius (\ref{eq:rescalHorizon}) and local  temperature distribution $T_{\text{eff}}(X,\theta)=1/\beta_{\text{eff}}(X,\theta)$ (\ref{eq:effLocalTemp}) as functions of $X$ for $q=4$. The un-deformed temperature is chosen to be $\beta=3$. The dotted lines illustrate the un-deformed horizon radius, and the orange sold lines illustrate the re-scaled horizon radius (\ref{eq:rescalHorizon}). The outermost circles illustrate the cutoff surface with the distribution of the local effective temperature (\ref{eq:effLocalTemp}).  {In these plots, we take the cutoff to be $\tilde{r}_{c}=5+2\pi/\beta$, giving $\theta_{cut}\approx 0.44.$}}
\label{fig:TempDistVsHorizon1}
\end{center}
\end{figure}

For the re-scaled horizon radius \eqref{eq:rescalHorizon}, let us estimate a local effective (inverse) temperature, which may
be related to the effective temperature defined as  \eqref{eq:effLocalTemp}.
First, let us assume that the observer sitting at $X$, who measures the temperature, can not distinguish the standard BTZ metric with the ``constant horizon radius" $r'_{0}(X,\theta)$ from the metric we have seen in the above discussion. In other words, the typical scale of the observer is assumed to be comparable with the re-scaled horizon radius $r'_{0}(X,\theta)$. 
Then, a thermal periodicity for ``the observer sitting at $X$", for who the metric can be approximated by the standard BTZ metric with the constant horizon radius $r'_{0}(X,\theta)$, is given by the following inverse temperature, 
\begin{equation}
	\begin{aligned}
		\tilde{\beta}_{\text{eff},x=X;\theta} = \frac{2\pi}{ r'_{0}(X,\theta) }= \frac{2\pi}{ r_{0} }f(X,\theta)= \beta \, f(X,\theta),
	\end{aligned}
\end{equation}
where we defined the un-deformed black hole inverse temperature $\beta$ by
\begin{equation}
	\beta=\frac{2\pi}{ r_{0} }.
\end{equation}
{One can confirm it from the holographic one-point function which is determined by geodesics distance between the boundary point $X$ and the black hole.}
By considering other observers sitting at different points, we can extend the above result to other points. Therefore, by collecting their results, the temperatures for observers sitting at $X$ are upgraded to the local temperature,
\begin{equation}
	\tilde{\beta}_{\text{eff}}(X,\theta) \coloneqq \tilde{\beta}_{\text{eff},X;\theta}= \beta \, f(X,\theta).
\end{equation}
This is equal to the boundary result \eqref{eq:effLocalTemp} as expected. 
In Fig. \ref{fig:TempDistVsHorizon1}, we show the $\theta$ dependence of the local temperature with the horizon radius \eqref{eq:rescalHorizon} for $q=4$ and various $\theta$. The local effective temperature $T_{\text{eff}}=1/\tilde{\beta}_{\text{eff}}$ takes the maximal value at the points where the re-scaled horizon takes the maximal value, while that takes the minimum value at the points where the re-scaled horizon takes the minimum value.

\section{Discussions}\label{sec:discussions}
Here, we will discuss some interesting aspects of our findings in this paper.

{\bf SSD limit:} Through this paper, we did not take the SSD limit, where $\theta \rightarrow \infty$. 
This is because the physical quantities such as the entanglement entropy can be infinite. 
We will explain why they become infinite in the SSD limit.
In the SSD limit, the Hamiltonian densities at $x= \f{mL}{q}$ vanishes.
Consequently, at $x=\f{mL}{q}$, $e^{-\beta H_{\text{q-SSD}}}$ can not tame the UV divergence.
Therefore, the entanglement entropy for the intervals, including the fixed points, becomes divergent. 
On the gravity side, the black hole horizon there becomes infinite.

{\bf Physical picture:}
Now, we give an interpretation of the entanglement entropy in the effective high temperature and high temperature region with the large $\theta$.
This interpretation may describe the $\mathcal{O}(1/\epsilon)$ term of the entanglement entropy which is determined by the theory-dependent piece of the entanglement entropy.
The theory-dependent piece is given by the minimal surface ending at the constant surface, $r_{\text{UV}}$, along $r$ coordinate, not $r'$ defined in Section \ref{sec:gravitydual}. 
Here, we assume $r_{\text{UV}} \gg 1$.
At $r_{\text{UV}} \gg 1$, the bulk metric approximately reduces to
\be
ds^2_{\text{bulk}} \approx r^2_{\text{UV}}\left(dy_{\xi}^2+ d\tau^2\right),
\ee
where $dy_{\xi}$ is defined as $dy_{\xi}=(d\xi -d \overline{\xi})/2i$.
Thus, in the coordinates, $(y_{\xi}, \tau)$, the boundary metric reduces to the uniform one.
We assume that the entanglement entropy is determined by the number of the quasiparticles in the subsystem, and in $(y_{\xi}, \tau)$ these particles uniformly distribute.
We define the density of quasiparticles as
\be \label{eq:effective-EE}
\rho_{\text{Q.P}.}= \f{S_{\text{Thermal}}}{L_{\text{eff}}}=\f{c\pi}{3\beta}.
\ee
Furthermore, we define the effective subsystem size as 
\be
l^{A_i}_{\text{eff}}=\int_{y_{\xi}(X_2)}^{y_{\xi}(X_1)}dy_{\xi}=\f{L\cosh{2\theta}}{q\pi}\left[\arctan{\left(e^{2\theta}\tan{\left(\f{q\pi X_1}{L}\right)} \right)}-\arctan{\left(e^{2\theta}\tan{\left(\f{q\pi X_2}{L}\right)} \right)}\right] 
\ee
We assume that the term at $\mathcal{O}(1/\epsilon)$ is given by 
\be \label{eq:effective-EE}
S_{A_i}=\rho_{\text{Q.P.}} \times l^{A_i}_{\text{eff}}.
\ee
If we take the large $\theta$ limit, and then $l^{A_i}_{\text{eff}}$ reduces to
\be
\begin{split}
    l^{A_i}_{\text{eff}} \approx \begin{cases}
        \f{L}{2q\pi }\f{\sin{\left[\f{q\pi (X_1-X_2)}{L}\right]}}{\prod_{i=1,2}\sin{\left[\f{q\pi (X_i)}{L}\right]}}~&~\text{for}~i=1\\
        \f{l \cdot L}{2q}e^{2\theta}~&~\text{for}~i=2
    \end{cases}.
\end{split}
\ee
In this limit, (\ref{eq:effective-EE}) matches the entanglement entropies calculated in the twist operator formalism.

{\bf Bulk reconstruction and relative entropy:} For simplicity, we define the subsystem $\mathcal{V}$ as
\be
\mathcal{V}=\left\{x\bigg{|} \f{mL}{q}<x<\f{(m+1)L}{q}\right\}.
\ee
We consider the reduced density matrix associated with $\mathcal{V}$ for (\ref{eq:thermal-state-with-inhomogeneous}), and then define modular Hamiltonian for this subsystem as
\be
H_{\text{Mod}}:=-\f{\log{\rho_{\mathcal{V}}}}{\beta}.
\ee
As in (\ref{eq:approximation-of-H}), $H_{\text{q-M\"obius}}$ factorizes into ${H_{\text{SSD},L/q,m}}$, which acts on the spatial interval between $mL/q$ and $(m+1)L/q$.
Therefore, in this limit, the modular Hamiltonian may be approximated as 
\be
H_{\text{Mod}} \approx {H_{\text{SSD},L/q,m}}.
\ee
By using this modular Hamiltonian in the regime, where this approximation is valid, it would be interesting to discuss the bulk reconstruction \cite{2016JHEP...06..004J} and relative entropy \cite{Sarosi:2017rsq}.


{\bf Entanglement transition induced by the curvature:} In this paper, we explored the thermodynamical and entanglement properties of the system in $2$d holographic CFTs on the curved spacetime, and then we found that this system exhibits the thermal and entanglement phase transitions induced by the spacetime. 
In 2d holographic CFTs, we took the large central charge limit.
At the leading order in this limit, the thermal and entanglement entropies exhibit these phase transitions. 
We expect the systems in $2$d free CFTs to exhibit the crossover of the thermal and entanglement entropies.
Therefore, it would be interesting to find other systems exhibiting these phase transitions induced by the spacetime even without the large central charge limit.
This may lead to the realization of the experimental system  exhibiting these phase transitions induced by the spacetime.

{\bf Quantum correlation of 2d Holographic CFTs on the curved spacetime:} In this paper, we investigated the mutual information, which measures both quantum and classical correlations, on the curved spacetime, and we found that the characteristic length scale of the mutual information has position dependence, such as shown in \eqref{eq:correlaLengthSuperHigh}, in some parameter regimes. This originates from the curved nature of the background spacetime. We expect that other correlation measures also have this sort of position dependence induced by the curved background. For example, it would be interesting to investigate quantum correlation measures, such as logarithmic negativity, odd entropy, and reflected entropy, which measure only quantum correlation, on the curved spacetime.

{\bf Entanglement dynamics on a cosine-square (CS) deformed background:}
In this paper, we focused on the case $\theta\geq 0$. However, the scenario where $\theta$ is negative is also of interest. In particular, the $\theta\to -\infty$ limit corresponds to the cosine-square deformation (CSD) limit. For this case, the stable and unstable fixed points for positive $\theta$ become the unstable and stable fixed points for negative $\theta$, respectively. Noting this correspondence, we can easily extend our discussion to the negative $\theta$ case.

\section*{Acknowledgement}
We thank Tadashi Takayanagi, Song He, Huajia Wang, Cheng Peng, and Chen Bai for useful discussions. 
M.N.~is supported by funds from the University of Chinese Academy of Sciences (UCAS), funds from the Kavli
Institute for Theoretical Sciences (KITS).
K.T.~is supported by JSPS KAKENHI Grant No.~21K13920 and MEXT KAKENHI Grant No.~24H00972. 
M.W.~is supported by Grant-in-Aid for JSPS Fellows No.~22KJ1777 and by MEXT KAKENHI Grant No.~24H00957.
\appendix

\bibliographystyle{ieeetr}
\bibliography{reference}
\end{document}